\documentclass[11pt,pra]{revtex4}

\usepackage{amssymb}

\usepackage{times}
\usepackage{verbatim}
\usepackage{graphicx}
\usepackage{amsmath}
\usepackage{subfigure}
\usepackage{setspace}
\usepackage{wrapfig}
\usepackage{sidecap}
\usepackage{bm}
\usepackage{color}
\usepackage{setspace}

\begin{document}

 \title{Ultracold Fermi Gases with Emergent SU(N) Symmetry}

\author{Miguel A. Cazalilla}
\affiliation{Department of Physics, National Tsing Hua University, and National Center for Theoretical Sciences, Hsinchu City, Taiwan},
\author{Ana Maria Rey}
\affiliation{NIST, JILA and Department of Physics, University of Colorado, Boulder, US}

\begin{abstract}
 We review  recent experimental and theoretical progress on ultracold alkaline-earth Fermi gases with emergent SU$(N)$ symmetry. Emphasis is placed on describing the ground-breaking
experimental achievements of recent years. The latter
include the cooling to below quantum degeneracy of various isotopes of ytterbium and strontium,  the demonstration of optical Feshbach resonances and the optical Stern-Gerlach effect, the realization of a Mott insulator of $^{173}$Yb atoms, the creation of various kinds of Fermi-Bose mixtures and the observation of many-body physics in  optical lattice clocks. On the theory side, we survey the zoo of phases that have been predicted for both gases in a trap and loaded into an optical lattice, focusing on two and three-dimensional systems. We also discuss some of the challenges that lie ahead for the  realization of such phases, such as reaching the temperature scale required to observe magnetic and more exotic quantum orders, and dealing with collisional relaxation of excited electronic levels.
\end{abstract}

\date{\today }
\maketitle

\tableofcontents

\section{Introduction}

Prior to the late 20th century, matter was primarily something to
be probed, dissected, and understood.  Now, in the early years of the
21st century, matter is something to be synthesized, organized, and
exploited for broader purposes, both at the level of basic research
and for numerous technological applications. One emerging area of research in this century  is to ultimately implement Richard Feynman's pioneering ideas of quantum simulation \cite{Feynman1982} and quantum information \cite{Feynmanbook}. We want to design in the laboratory artificial, fully controllable  quantum systems, and use them to  mimic models of many-body systems relevant for  otherwise intractable problems in materials physics and other branches of modern quantum Science.

 In fact, recent advances in cooling and trapping alkali atoms has brought us closer to realizing Feynman's dreams.
Their simple  electronic structure (they possess a single valence electron) has allowed  a clean characterization of their hyperfine levels,  greatly facilitating the development of extremely effective  trapping  and quantum control techniques.
Using these atoms,  major breakthroughs have been achieved such as a detailed
understanding of the BEC to BCS crossover~\cite{Giorgini2008,Bloch2008r} and the implementation of both Fermi and Bose Hubbard models \cite{Bloch2008r,Jaksch1998,Greiner2002a,Jordens2008,Schneider2008}. 

 Nevertheless, the inherent  ``simplicity'' of alkali atoms
  introduces major limitations to the phenomena that can be explored with them. For example, the actual observation of quantum
magnetism in the Fermi/Bose Hubbard models has been hindered by the low entropy requirements set by the energy scales  of the effective spin-spin interactions.   
In this regard, systems exhibiting more complex internal structure
could be an excellent platform for exploring a wider range of many-body  phenomena.
They also hold the promise  of the discovery of new states of matter that go beyond the possibilities already offered by conventional condensed matter systems. During the last few decades, there have been  exciting  advances  in this direction, as new capabilities for cooling, trapping, and manipulating  more complex systems such as trapped ions, magnetic atoms, Rydberg atoms, polar molecules, and alkaline-earth atoms have been demonstrated. Here we concentrate our attention on alkaline earth atoms.

Strictly speaking, alkaline-earth atoms (AEA)  lie in group-II of the periodic table. However, we will also include  others with similar electronic structure like Ytterbium (Yb). These atoms   have unique atomic properties which make them ideal  for the realization of ultra-precise atomic clocks. Lately, as we shall explain below, they are also attaching a great deal of attention for their interesting many-body physics and the possibilities that they offer for the quantum simulation of complex quantum systems.

\begin{figure}{}
\centering
    \includegraphics[width=80mm]{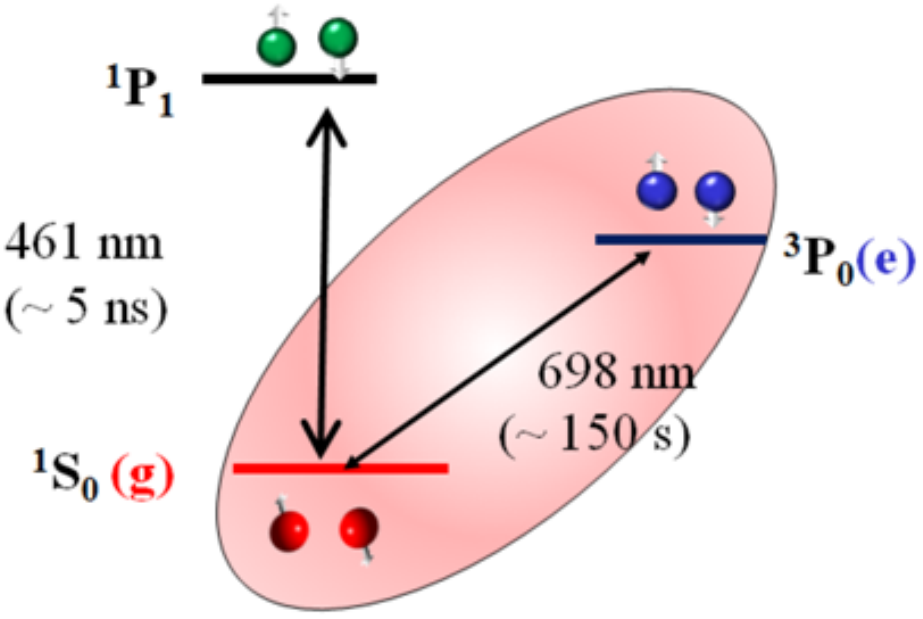}
\caption{  Energy levels of ${}^{87}$Sr. The singlet,$^1S_0$, and  the triplet,  $^3 P_0$, states have an inter-combination line as narrow as a few mHz.}
\label{levels}
\end{figure}

 Nevertheless, before immersing ourselves in the study of their fascinating many-body physics, it is worth recalling that atomic clocks provide one of the most striking illustrations  of the unique advantage of AEA over alkali atoms. State-of-the-art  optical  atomic clocks  use  fermionic AEA, such as Sr or Yb~\cite{Derevianko2011}. Those clocks have already  surpassed the accuracy of the Cs
 standard~\cite{Ludlow2008}. The most stable of these clocks now operate near the quantum noise
 limit~\cite{Nicholson2012,Ludlow2013} and just recently, thanks to  advances in modern precision laser spectroscopy,  are becoming the most precise  in the world, even surpassing the accuracy of  single ion standards~\cite{Bloom2013}.   The stability of the neutral atom optical clocks arises from the extremely long lived singlet, $^1S_0$, and triplet states $^3 P_0$,  generally referred to as clock states, with intercombination lines both electric and magnetic dipole
forbidden and as narrow as a few mHz--- nine orders of magnitude lower
than a typical dipole-allowed electronic transition (See Fig.~\ref{levels}). It is impossible to
achieve this level of clock stability with conventional alkali atoms,  due to the decoherence that arises from the intrinsic sensitivity of the hyperfine ground states to magnetic field fluctuations and/or to intensity and phase noise on the optical fields.

 Returning to many-body physics and quantum emulation using AEA, in this article we attempt to review the  experimental and theoretical progress in this area. Given the large amount of recent research, we mainly focus on the consequences of their emergent SU$(N)$ symmetry of the AEA Fermi gases. Thus, we have tried to capture  ``snapshots'' of the ongoing experimental progress. As far as theory is concerned, we also have attempted to provide a survey of some of the most important and interesting theoretical proposals. Therefore, our selection of topics in the latter regard is rather subjective, and the emphasis has been  placed on providing a pedagogical introduction to some of the subjects rather than on providing an exhaustive survey of the available literature. As a consequence, some topics have been left out. For instance,
the application of AEA for quantum information purposes will not be discussed here and we refer the interested  reader to Ref.~\cite{Daley2011r}. Another topic that we do not touch in depth is the physics of one-dimensional (1D) systems. This subject has been the focus of theoretical interest in recent years, especially concerning quantum magnetism in 1D lattice systems  (see e.g.~ Refs.\cite{Nonne2013,Nonne2010,Szirmai2013,Szirmai2008} and references therein).  For trapped systems on the continuum, we refer the interested reader to the excellent recent review article on this subject by Guan
\emph{et al.}~\cite{Guan2013} and point out that just recently the first experimental exploration of  the fascinating role of SU$(N)$ symmetry  in  an array of 1D fermionic tubes has been reported in Ref.~\cite{Fallani2014}.

The outline of this article is as follows: We begin in section~\ref{sec:sun} with
a review of the work leading to the observation that AEA posses an emergent SU$(N)$ symmetry (for a brief review of the group theory relevant to SU$(N)$, see Appendix~\ref{app:sun}). Although this was a theoretical prediction, it was  based on a number
of experimental observations associated  to  the unique atomic structure of  AEA. The emergent SU$(N)$ symmetry has not only
important  consequences in atomic molecular and optical systems and condensed matter physics, many of them reviewed here,
but also in other fields in physics as well. In sections~\ref{sec:exptrap} and \ref{sec:oplatt}, we review
the experiments that have been performed so far both in traps and in optical lattices, respectively.
The review of theoretical results begins in section~\ref{sec:flt}, where the theory of SU$(N)$
Fermi liquids and their instabilities, including the Bardeen-Cooper-Schrieffer (BCS) instability, are surveyed. Whereas the discussion in this section mainly  applies to gases in a trap, in section~\ref{sec:thoplat} we turn our attention to quantum phases that are intrinsic to lattice systems. Focusing on the deep lattice limit, we discuss both the Fermi Hubbard and Heisenberg models with SU$(N)$ symmetry. Finally, in section~\ref{sec:other}, we conclude by discussing other interesting models that can be engineered using alkaline-earth atoms. Appendix~\ref{app:sun} contains a brief summary of the most important mathematical results about the SU$(N)$ group and appendices~\ref{app:landau} and
\ref{app:ng} contain some technical details of the topics discussed in section~\ref{sec:flt}.

\section{Alkaline Earth Fermi gases: An emergent SU$(N)$ symmetry}\label{sec:sun}

Recently, it has been realized that AEA unique atomic structure has   fundamental features  which make them attractive for the study of many-body phenomena.
One of  their most appealing property   is an  emergent SU$(N)$ symmetry  in the nuclear spin degrees of freedom  \cite{Cazalilla2009,Gorshkov2010} and many  of the consequences arising from it  remain to be exploited and understood.


\subsection{Background and Precedents}\label{sec:backgrnd}

In order to understand how the SU$(N)$ symmetry emerges  at ultracold temperatures, let us first recall the pioneering work by Lee, Yang, and Huang~\cite{LeeHuangYang1957}. These authors considered the thermodynamic description of interacting gases well below their quantum degeneracy temperature and
argued that,  provided the range of
the interactions is much shorter than the inter-particle distance ({\it i.e.} the gas is ``dilute''), the complicated inter-atomic potentials are
well approximated by the pseudopotential $V(\boldsymbol{r}) = \frac{2\pi \hbar^2 a_2}{\mu} \delta(\boldsymbol{r}) \partial_{r} \left[ r  \cdot\right]$, where $r = |\boldsymbol{r}|$ is the relative separation of the colliding particles, $\mu$ their  reduced mass ($=$ half the bare mass for identical particles), and
 $a_s = - \lim_{k\to 0} \delta_s(k)/k$ the scattering length ($\delta_s(k)$ is the $s$-wave scattering phase-shift). The latter
is the only parameter needed to characterize the interactions, since  at  ultra-low temperatures higher partial waves are suppressed by the centrifugal barrier.

  As formulated by Lee, Huang, and Yang, the pseudo-potential applies to bosons and spin-$\frac{1}{2}$ fermions only.
It has been later noted by Yip and Ho~\cite{YipHo1999} that for spin$-F$ fermions, this pseudo-potential must be generalized to:
\begin{equation}
V(\boldsymbol{r}) = \sum_{\mathrm{even}\: j=0}^{2F-1} \frac{2\pi \hbar^2 a^j_{s}}{\mu}  \,  \delta(\boldsymbol{r}) \partial_{r} \left[ r  \cdot\right]\, \mathcal{P}_{j}, \label{eq:yipho}
\end{equation}
where $\mathcal{P}_{j}$ is the projector onto states with total spin equal to $j = 0, 2, \ldots, 2F-1$. Only the even  $F$ values can interact via $s$-wave collisions since due to quantum statistics those are the only ones that have
an  associated  spatial  wavefunction which  is  anti-symmetric.  Hence, it follows that $2F-1$ scattering lengths are needed to describe the interaction between spin-$F$ fermions. Crudely speaking, the differences between the scattering lengths $a^0_{s},\ldots,
a^{2F-1}_{s}$ stem from the different configurations the electronic shell of the
colliding atoms can adopt for the possible values of $F$.  In the presence of a   large magnetic field, $F$ is not longer a good quantum number and the scattering lengths between  states with different projection along the quantization  direction can  become also  different \cite{Chin2010}.

 However, Eq.~\eqref{eq:yipho} can exhibit a much larger  symmetry than  naively expected for a higher spin representation of SU$(2)$. As Wu and coworkers~\cite{Wu2003} noticed for $F = \frac{3}{2}$, the four-component spinor representation of SU$(2)$ is isomorphic to a spinor representation of the SO$(5)$ group without fine tuning. These authors also pointed out that, for the $F=\frac{3}{2}$ members of the AEA family
$^{135}$Ba and $^{137}$Ba, the scattering lengths $a^2_{s}$ and $a^0_{s}$ should
have similar values due to the completely filled outer electronic shell of Barium. These atoms were thus located  close to the  SU$(4)$
symmetric line in the  phase diagram of Ref.~\cite{Wu2003}.

Alkali gases with  approximate SU$(3)$ symmetries have been considered by a number
of authors, beginning with the pioneering work by Modawi and Leggett~\cite{ModawiLeggett1997}, who studied  BCS pairing in a quantum degenerate mixture of the three spin-polarized hyperfine states of  $^{6}$Li.  Honerkamp and Hofstetter~\cite{Honerkamp2004,Honerkamp2004b} also considered this system as well as mixtures of $N$ hyperfine states of $^{40}$K . The three-component $^{6}$Li system has been recently realized experimentally~\cite{Jochim2008,Ohara2009} and  evidence of the emergence of a SU(3) symmetry at large magnetic  fields (at which the  electronic  and   nuclear spin degrees of freedom start to become decoupled )
 has been reported. However at  moderate magnetic fields the SU$(3)$ symmetry breaks down.

\subsection{Alkaline-earth and Ytterbium Atomic (AEA) Gases}

 For the AEA in the ground state ($^1S_0$), the electronic degrees of freedom have neither spin nor orbital angular momentum ($J=0$). The nuclear spin ($I>0$), present only in the fermionic isotopes, is thus decoupled from the electronic state  due to the absence of hyperfine interactions.  Note  that all bosonic AEA have $I=0$ due to their even-even nuclei configuration. Interestingly, the excited state  $^3 P_0$ also has, to leading order, vanishing  hyperfine interactions and hence highly decoupled  nuclear and electronic spins \cite{Boyd2006}.

 The electronic-nuclear spin decoupling in the fermionic isotopes not only  allows for an independent manipulation of  their nuclear and electronic  degrees of freedom,
 but also imposes the  condition  that the scattering parameters  involving the ${}^1S_0$ and ${}^3P_0$  states  should be  independent of the nuclear spin, aside from the
restrictions imposed by fermionic antisymmetry. Thus, with great accuracy (see
discussion below), in the clock states  all the scattering lengths are equal, i.e. $a^{j}_s = a_s$ (for $j=0,2,\ldots, 2F-1$). Under these  conditions the interaction  and kinetic Hamiltonians  become
SU(N) spin symmetry
(where $N = 2 I
+ 1= 2F+1$)~\cite{Cazalilla2009,Gorshkov2010}.

 For the ${}^1S_0$ it has been theoretically determined that the variation of the scattering length for the various nuclear spin components, should be smaller $\delta a_s/a_s\sim 10^{-9}$ \cite{Gorshkov2010}. Although for the  $^3 P_0$ electronic state, the
decoupling is slightly broken by the admixture with higher-lying $P$
states with $J\neq 0$, this admixture is very small and the resulting
nuclear-spin-dependent variation of the scattering lengths is also
expected to be very small, of the order of $10^{-3}$ .

  The bounds on the variation of the scattering lengths, $\delta a_s/a_s$  associated to the various nuclear spin projections  are based on the fact that the scattering length is just a measure of the semiclassical phase, $\Phi$,  accumulated by the colliding atoms from the turning point to infinity (computed at zero energy) \cite{Gorshkov2010}.  The variation of the phase, proportional to is thus proportional to  $\delta a_s/a_s\sim \delta \Phi=\delta V \Delta t/\hbar$, with $\Delta t\sim1$ps the total time in the short-range part of the collision and $\delta V$ the typical energy difference associated with different nuclear spin projections during this time. For the  ${}^1S_0$ state, the latter can be estimated using second order perturbation theory as $\delta V/h\sim E_{hf}^2/(E_{opt}h)\sim  200$ Hz, where $E_{hf}/h\sim  300$ MHz is the approximate value for the hyperfine splittings in ${}^3 P_1$ and $E_{opt}/h\sim  400$THz is the optical energy difference between ${}^1S_0$ and ${}^3 P_1$. This leads to the $10^{-9}$ estimate. For the ${}^3P_0$, the second order formula might be incorrect since, the associated  molecule states separated by the fine structure energy at large distance  may come orders of magnitude closer at short range. Thus to assume $\delta V\sim E_{hf}$ accordingly to  first order perturbation theory is a more realistic and conservative estimate, which yields  $\delta \Phi\sim 10^{-3}$.

\subsection{Relevance of SU$(N)$ symmetry }

It cannot go unnoticed that the availability of fermionic systems exhibiting
an enlarged SU$(N)$ symmetry with  $N$ as large $10$ can be  interesting for other fields of physics beyond research on ultracold gases. For instance, in particle physics the theory of quantum chromodynamics (QCD) --which currently provides  the most fundamental description of the atomic nucleus and the nuclear interactions-- contains two kinds of SU$(3)$ groups. A  global flavor SU$(3)$ symmetry group, whose discovery won the Nobel prize for Gell-Mann, and the gauged color SU$(3)$. The latter  describes the origin of the forces that confine the quarks inside baryons and mesons through the  exchange of SU$(3)$ gauge bosons known as gluons. In the field of nuclear physics, the SU$(6)$ group has also been considered as candidate to unify the
description of baryons and mesons into a single group capable of accounting for both the flavor SU$(3)$ and spin SU$(2)$ symmetries~\cite{GurseyRadicati1964}. Indeed, the interesting analogies between ultracold gases with enlarged SU$(N)$ symmetry and cold dense QCD Matter have been already noticed by several authors (see e.g.~\cite{He2006,Rapp2007,Cazalilla2009,Ozawa2010} and references therein).

The SU$(N)$ symmetry can also have remarkable consequences in  quantum many-body systems. For example in a  SU(2) antiferromagnet, which characterize for example spin $1/2$ particles with spin rotation symmetry,
every pair of spins minimizes its energy by forming a singlet. The same spin, nevertheless, can participate in only one singlet pair with one of its neighbors. In principle, this constrain can generate geometrical frustration and prevent magnetic ordering. However,
spin-$\frac{1}{2}$  particles  tend to find a compromise and   often  become magnetically ordered  with decreasing temperature. A typical example of  that compromise is found in the SU(2) Heisenberg model on a triangular lattices where, in the ground state, adjacent spins anti-align at $120^\circ$ degrees.

  On the other hand, systems with an enlarged number of degrees of freedom,  and exhibiting  SU$(N> 2)$ spin rotation symmetry, suffer from massive degeneracies. The latter tend to  favor absence of magnetic ordering even classically~\cite{Hermele2009}. Quantum mechanically, this translates into ground states containing massive spin superpositions that give rise to  topological order and long range quantum
entanglement~\cite{Kalmeyer1987,Wen1989}. Examples of long range quantum entanglement  states are fractional quantum Hall states and the ground state of Kitaev's toric code \cite{Kitaev2006}.

Indeed, the identification of the SU$(N)$ symmetry  as a unique resource for dealing with unconventional magnetic states has  a long history in condensed matter physics~\cite{Read1983,Coleman1983,Read1987r,Affleck1988,Marston1989,Read1989}.  A generalization of the symmetry from SU$(2)$ to SU$(N)$ introduces a perturbative parameter, namely $1/N$. A large $N$ expansion is particularly useful when dealing with problems of quantum magnetism  for which there is no other  small parameter that allows
for a perturbative treatment. The Kondo  impurity problem, the Kondo lattice model~\cite{Read1983,Coleman1983,Read1987r}
and the Hubbard
model~\cite{Read1989,Affleck1988,Marston1989}  are some examples of systems where a large $N$ expansion has been shown to be useful. In such systems, fluctuations about the mean field solutions appear at order $\frac{1}{N}$. The  hope is
that even at finite $N$, the  $\frac{1}{N}$ corrections can remain  irrelevant and   thus the mean field solution a good approximation.
However, in this context, the enlarged SU$(N)$ symmetry has been often regarded as a mere mathematical construction  without a real physical motivation~\footnote{An exception is the Kondo problem discussed in Refs.~\cite{Read1983,Coleman1983,Read1987r}, for which $N$ corresponds to the number of possible degenerate magnetic configurations of the impurity atom.} and in many cases just as a means to develop approximate solutions for SU(2) systems. The observation that SU$(N)$ symmetry  naturally emerges in  the nuclear spin degrees of freedom of fermionic alkaline earth atoms thus  raises the exciting potential opportunity of
bringing  it back to reality and opens the possibility of exploiting its remarkable consequences for the first time in the  laboratory.

Finally, it is also worth mentioning that enlarged unitary symmetries have been also be used  in other problems in
solid state physics, such as the quantum Hall effect in multi-valley semiconductors~\cite{Arovas1999,Ezawa2003}. In such a systems,
the massive degeneracy of the Landau levels is supplemented by a large degeneracy in spin and valley-spin, which favors ferromagnetic states and complex spin-valley textures~\cite{Arovas1999,Ezawa2003}.
 A recent revival of the interest in these systems has been brought
 by graphene~\cite{CastroNeto2009}, which can be regarded as a two-valley zero-gap  semiconductor. Graphene exhibits a
SU$(4)$ spin-valley symmetry~\cite{CastroNeto2009,Goerbig2011}, which, although  weakly broken by the long-range part of the Coulomb interaction, plays an important role in determining
the properties of the ground state both in the
integer~\cite{Kharitonov2012,Kharitonov2012b}
and fractional quantum Hall effect~\cite{Goerbig2007,Goerbig2011}. Finding connections
between these phenomena and  the many-body physics
of AEA remains an interesting challenge for both
experimentalists and theorists.



\section{Experiments with Trapped Ultracold Gases}\label{sec:exptrap}

Owing to their unique properties, recently, substantial experimental efforts have been directed at cooling, trapping, and manipulating AEA and many of the capabilities previously demonstrated with alkaline atoms are starting to be reproduced
with AEA. These include laser cooling  down to microKelvin temperatures, trapping  in optical potentials
for several seconds, evaporative cooling to quantum degeneracy, the  demonstration of a high degree of control over both
internal and external degrees of freedom, imaging and resolving  the various hyperfine components  using optical Stern-Gerlach,  demonstrating control   of interaction parameters via optical and magnetic Feshbach resonance, and the realization of a Mott Insulator. In  this section, we first present a summary of those experimental developments for trapped gases. In the following section, we review the experiments dealing with Fermi gases on optical lattices.

\subsection{Experimental determination of the scattering length}

A  natural manifestation of the   SU$(N)$ symmetry is the conservation of each of the nuclear spin components during a collision. This is in stark
contrast to the  smaller SU(2) symmetry exhibited by  alkali atoms  which allows for spin changing processes for $F > \frac{1}{2}$. This is because, as described in section~\ref{sec:backgrnd}, for the latter   the scattering lengths depend on the total $F= J+I$ of the colliding atoms an therefore, during collisions  the bare nuclear spin degrees of freedom get effectively mixed. Thus, even though the total spin magnetization ($m_F$) is always conserved, it is possible to have  spin changing collisions, for example between two  atoms with angular momentum projection  $m_F=0$, into one in $m_F=1$ and the other in $m_F=-1$.

 As emphasized above, although the $s$-wave scattering length is the only parameter that fully characterizes the collisional properties of ultra-cold gases, it  is very sensitive to the ground state interatomic
potential, an thus naive ab initio calculations in general fail to determine it \cite{Gribakin2003}.  Consequently, we need to rely on experiments for its actual determination. Among those experiments, we can mention: cross-dimensional rethermalization measurements, time of flight images, and one and two-color photo-association spectroscopy.

\begin{figure}{b}
\centering
    \includegraphics[width=110mm]{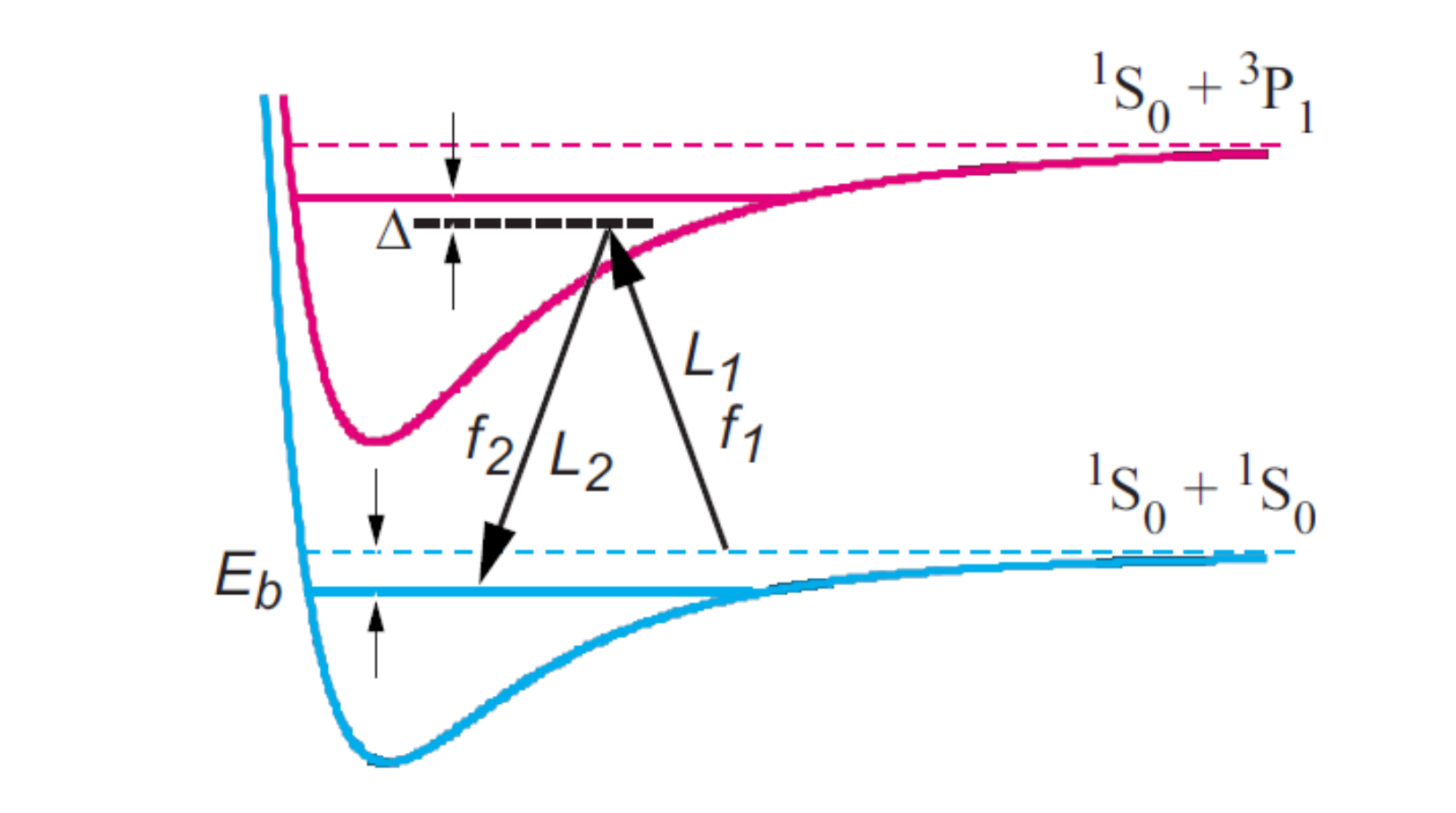}
    \caption{(Reproduced from Ref. \cite{TwoPAYb}) Schematic description of the two-color
PA spectroscopy. The laser $L_1$ drives the one-color PA transition.
The laser $L_2$ couples the bound state in the excited molecular potential
to the one in the ground molecular potential. The detuning 	
of the PA laser with respect to the one-color PA resonance is set to several MHz for the Raman spectroscopy, while
is set to zero for the Autler-Townes spectroscopy}
\label{TCPA}
\end{figure}

Two-color photo-association (TPA) uses two laser beams to measure the binding energy of the weakly bound states of a molecular system~\cite{Kitagawa2008} (See Fig.~\ref{TCPA}). One, $L_1$, which probes a transition between  a pair of colliding ground state atoms and a  excited molecular bound state, and  a second, $L_2$,
which probes the bound-bound transition between the excited molecular bound state and  a ground molecular bound-state close to the dissociation threshold.  If $L_2$ is close to  resonance to the bound-bound transition, it causes the so called Autler-Townes doublet \cite{Autler1955}  and when the laser $L_1$ is also on resonance to the free-bound  transition, the atomic loss coming from the population of the molecular excited state  is suppressed due to quantum interference (Autler-Townes spectroscopy).  On the contrary,  if both lasers are off-resonant  and the frequency difference matches the binding energy of the ground molecular state, the lasers drive a stimulated Raman transition  from the colliding atom pair to the molecular ground state which can be detected as an atom loss (Raman spectroscopy).

For  AEA two-color photo-association (TPA) has become the most reliable and precise way to determine ground state scattering lengths. This is because  the absence of hyperfine structure in   the ${}^0S_1$  state (with no electronic orbital and spin angular momenta)  gives rise to a simple isotope-independent ground state molecular potential. The number of bound states in the molecular potential, which can be cleanly extracted from TPA can then  be used as an input parameter in a semiclassical theory \cite{Gribakin2003} which, together with  mass scaling,  can  determine the scattering length of all isotopes with    unprecedented precision.

In Table  \ref{table1} we display the measured values of the $s$-wave scattering lengths for various fermionic AEA along with other relevant data, such as  their mass, nuclear spins and emergent SU$(N)$ symmetries.
Note that the scattering length can vary from large negative to large positive values. The magnitude of the $s$-wave scattering length determines the feasibility of reaching quantum degeneracy for the various isotopes via evaporative cooling methods.

\begin{table}[t]
\begin{center}
\begin{tabular}{c|c|c|c|c}
\hline\hline
Atom Species & Mass (u) & Nuclear Spin ($I$) & Symmetry Group & Scattering Length (nm)\\
\hline\hline
$^{171}$Yb & $170.93$ & $\frac{1}{2}$ & SU$(2)$ & $-0.15(19)$ [$^{171}$Yb], $-30.6(3.2)$ [$^{173}$Yb] \cite{TwoPAYb} \\
$^{173}$Yb & $172.94$ & $\frac{5}{2}$ & SU$(6)$ & $10.55(11)$ [$^{173}$Yb], $-30.6(3.2)$ [$^{171}$Yb] \cite{TwoPAYb} \\
$^{87}$Sr & $86.91$ & $\frac{9}{2}$ & SU$(10)$ & $5.09(10)$ [$^{87}$Sr] \cite{PASr4}\\ 
\end{tabular}
\end{center}
\caption{Table of fermionic alkaline-earth atom (AEA) characteristics. The data is for the AEA species that have been so far cooled down to quantum degeneracy.}
\label{table1}
\end{table}


\subsection{Towards  a quantum degenerate gas}

\subsubsection{Ytterbium}

 The quest of achieving a quantum degenerate gas with group-II atoms started with Yb. Yb has five stable bosonic isotopes
$^{168,170,172,174,176}$Yb and  two fermionic isotopes, $^{171}$Yb with $I=1/2$ and $^{173}$Yb
with $I=5/2$.

The first experimental realization of a Bose Einstein Condensate (BEC) of ${}^{174}$Yb was reported in 2003  by the Kyoto group~\cite{unoYb}. The lack of hyperfine structure in the ground state of bosonic AEA prevents the use  of the
conventional magnetic trap for  BEC production  and evaporative cooling by a radio frequency knife. Instead, all-optical trapping and cooling methods are needed. Four years later, in 2007 all-optical formation of degenerate
fermionic ${}^{173}$Yb gas was achieved by the  Kyoto group by performing evaporative cooling of the six-nuclear spin-state mixture
in the optical dipole trap \cite{Tresyb}. Following it, a BEC of  ${}^{170}$Yb \cite{dosYb} and   ${}^{176}$Yb \cite{CuatroYb} were reported by the same group. The latter required  sympathetic cooling of  ${}^{176}$Yb with ${}^{174}$Yb, due to the fact that ${}^{176}$Yb has a negative scattering length. A
rapid atom loss in ${}^{176}$Yb atoms seen after cooling down the gas below the
transition temperature was  consistent with the expected  collapse of a ${}^{176}$Yb condensate due to attractive interactions.

The  ${}^{171}$Yb  fermionic isotope has a very small scattering length in its ground state, $a_s \approx -0.15$ nm, which prevents cooling by direct evaporation. However, in 2010 it was cooled to quantum degeneracy  via sympathetic  cooling with ${}^{176}$Yb .  This allowed to realize, in the presence of ${}^{173}$Yb,  the first SU(2)$\times$SU(6)  mixture  in ultra-cold gases \cite{Taie2010}.  Finally, despite the low natural abundance of   ${}^{168}$Yb, of the order of $0.13\%$, a BEC of this rare atomic species was obtained by direct evaporative cooling in 2011~\cite{CincoYb}. Thus, except from ${}^{172}$Yb, which is unstable to three body losses due to its large negative scattering length, quantum degenerate gases and/or mixtures of all the stable Yb isotopes have been produced by  the Kyoto group. Recently the creation of quantum degenerate gases of Ytterbium has been also reported by Sengstock's group in Hamburg~\cite{SeisYb}. The  production of  quantum degenerate mixtures of fermionic alkali-metal ${}^6$Li and bosonic Yb \cite{YbLimix,YbLimixtures} and fermionic Yb \cite{YbLimixtures}  has also been reported.

\subsubsection{Calcium}

Calcium was the first AEA, truly belonging to the
group-II elements of the periodic table, which was cooled down to quantum degeneracy. In 2009 at PTB (Germany) a BEC of ${}^{40}$Ca was reported \cite{UnoCa}. This was achieved in spite of the inelastic collisions associated with its large ground state $s$-wave scattering length ($18{\rm nm}<a_s<37{\rm nm}$)~\cite{PASr4}, by using a large volume  optical trap for initial cooling. A second Calcium
BEC  was reported in 2012 by  Hemmerich's group in  Hamburg~\cite{Dosca}. So far, no fermionic isotopes of Ca have
have been cooled below the quantum degeneracy temperature.

\subsubsection{Strontium}

 Strontium has three relatively abundant isotopes.  Two of them  are bosonic ${}^{86}$Sr and ${}^{88}$Sr  with relative abundance $\approx 9.9\%$ and  $\approx 82.6\%$ respectively  and one fermionic  ${}^{87}$Sr  with $\approx 7.0 \%$ and a nuclear spin $I=9/2$.

Initial efforts to reach quantum degeneracy with Sr gases failed due the  unfavorable scattering properties of the bosonic isotopes~\cite{Zero0Sr,ZeroSr}.
While the  scattering
length of  ${}^{88}$Sr is close to zero, the scattering length of the  ${}^{86}$Sr  isotope is
very large, $a_s \approx 40$ nm, leading to large detrimental loss of atoms by three-body recombination. The breakthrough for reaching quantum degeneracy came from the development of an efficient loading scheme which allowed to overcome the  low  natural abundance of  ${}^{84}$Sr (only $\approx 0.56\%$) and to take advantage of its favorable scattering length, $a_s \approx 6.5$ nm. A BEC of ${}^{84}$Sr was almost simultaneously reported by  two groups, Schreck's group (Insbruck)~\cite{UnoSr} and Killian's group (Rice University at Texas) \cite{DosSr}. This achievement was followed up by the cooling to quantum degeneracy of a spin polarized gas  ${}^{87}$Sr in  thermal contact with a BEC of ${}^{84}$Sr \cite{CuatroSr}  and  the corresponding mixture \cite{TresSr} by the same groups respectively.
 A  BEC of ${}^{86}$Sr was finally created, in despite of its large scattering length, by the  Innsbruck  group. This was achieved  by reducing the density
 in a large volume optical dipole trap \cite{CincoSr,Stellmer2013a}. Furthremore, a BEC of ${}^{88}$Sr  was produced by the Rice group, which used sympathetic cooling with ${}^{87}$Sr \cite{SieteSr}.  Finally, the quest of developing faster and better  pure optical methods towards reaching larger and colder samples  of AEA has recently lead to the implementation of a method based on laser cooling as the only cooling mechanism \cite{Stellmer2013b}.

\subsection{Control of Interactions: Optical Feshbach resonances}\label{sec:optfesh}

The ability to tune interactions in ultracold alkali-metal atomic
gases using  magnetic Feshbach resonances (MFR) has been a crucial step  for the exploration of  few and  many-body
physics in these systems \cite{Chin2010}. MFR, however, can not be used to tune interactions in ground  state of AEA due to
the lack of magnetic electronic structure.

 However, tuning interatomic interactions via Optical Feshbach resonances (OFR) is a feasible route in AEA. In a OFR  a laser  tuned near a photoassociative resonance is   used
to  couple a pair of colliding atoms to a bound
molecular level in an excited electronic level. The coupling induces a Feshbach
resonance and modifies the scattering length of the two colliding
atoms. In Ref. \cite{Ciurylo2005} it was predicted that OFR could be ideally implemented in AEA using a transition
from the singlet ground state to a metastable triplet
level. The possibility of   tuning
the scattering length with significantly less induced losses was based on  the long lifetime of the excited molecular state and a relatively large
overlap integral between excited molecular and ground
collisional wavefunctions.

There has been already a few  experimental demonstrations of the use of OFR to modify interaction properties in AEA, although  significant atom loss  has always been observed. The modification of the photoassociation spectrum by an OFB  in a thermal gas of  ${}^{172}$Yb  was
reported in Ref.~\cite{OFRYb1}. An OFR laser pulse of a 1D optical lattice turned on for several microseconds was used in Ref.~\cite{OFRYb2}  to  modulate the
mean field energy in a ${}^{174}$Yb  BEC.  In  Ref.~\cite{OFRSr2} an OFR in  a thermal gas of   ${}^{88}$Sr
was used to  modify its  thermalization and loss rates. More recently an OFR  in Ref.~\cite{OFRSr1} was used to control  the collapse and expansion of a   ${}^{88}$Sr BEC by moderate  modifications of the  scattering length.
The  use of  more deeply bound excited molecular states to help the suppression of atom-light scattering and to reduce the  background two-body loss could enhance the utility  of OFR  in AEA and  efforts in that direction are currently taking place in various labs. One important point to highlight, nevertheless, is that the direct use of OFR to control scattering properties  can destroy the SU$(N)$ symmetry since the ground state is directly coupled to an excited state which does possess a hyperfine structure.

\subsection{Imaging and detection of nuclear spin components}

\begin{figure}{}
\centering
    \includegraphics[width=110mm]{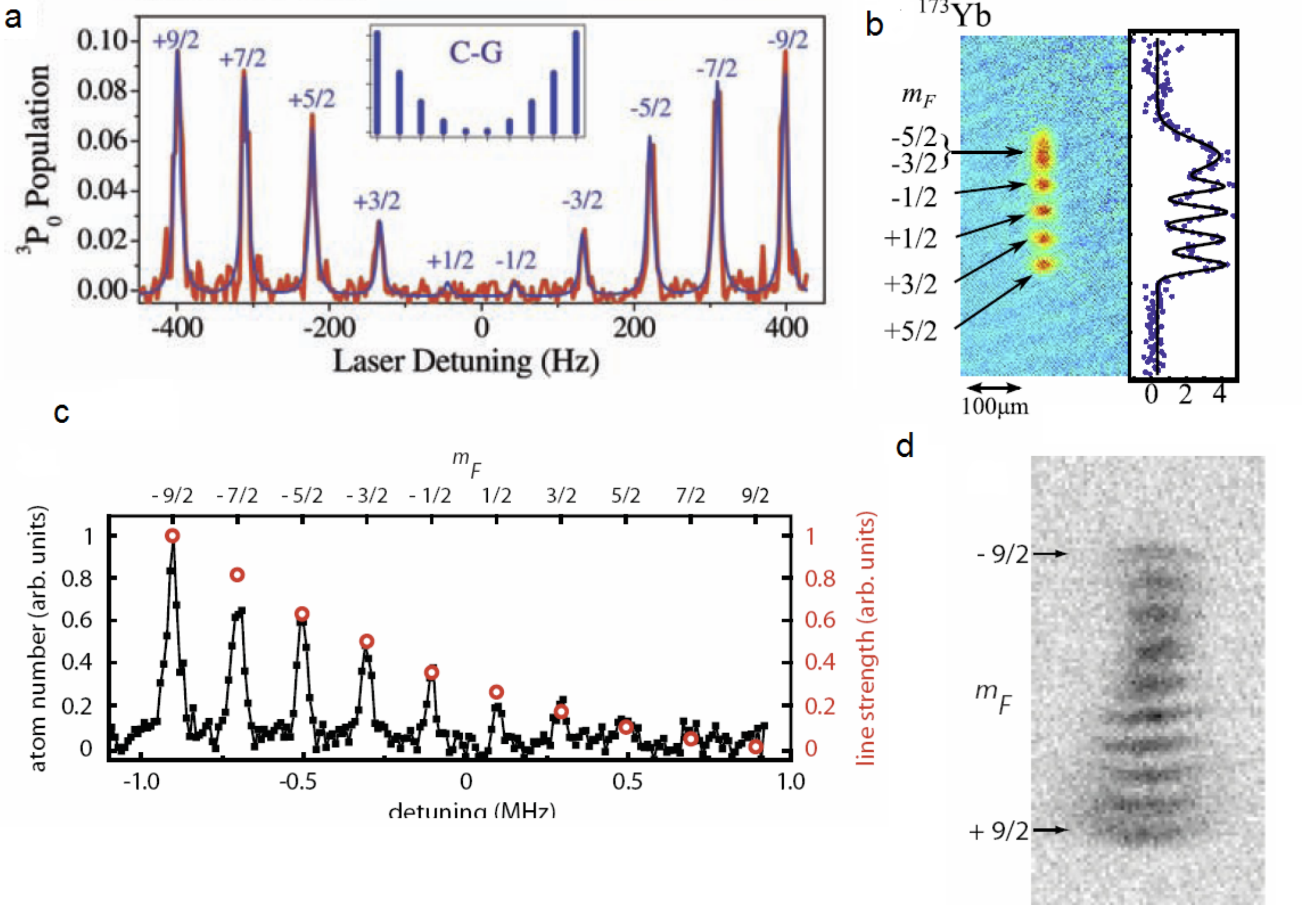}
    \caption{ Experimental resolution of the nuclear spin sublevels:  a) Spectroscopically interrogating the   ${}^1 S_0$- ${}^3 P_0$ transition in an optical lattice clock \cite{Boyd2006} operated with  ${}^{87}$Sr gas, b) in a quantum degenerate gas of ${}^{173}$Yb using Optical Stern-Gerlach (OSG) \cite{Taie2010}, c) Spectroscopically using the ${}^1 S_0$- ${}^3 P_1$ intercombination line  in a cold quantum  gas of ${}^{87}$Sr atoms at a temperature of $0.5 \mu$K \cite{Stellmer2011} and d) using Optical Stern-Gerlach (OSG)  in a quantum degenerate ${}^{87}$Sr gas \cite{Stellmer2011}.}
\label{nuclear}
\end{figure}

An important tool for probing AEA is the ability to separately resolve the different nuclear spin components (see Fig.~\ref{nuclear}).
In group-I elements like alkali atoms hyperfine states can be resolved and imaged by taking advantage of the well known  Stern-Gerlach technique. The latter uses the spin-state dependent force generated by a  magnetic-field gradient  to spatially split an expanding atom cloud in clouds of different  hyperfine levels.
However, this method cannot be used for AEA in the clock states for which $J=0$, due to their small magnetic moment which entirely stems from the nuclear spin. Let us recall that the nuclear magnetic mangeton is about three orders of magnitude smaller than the Bohr magneton and therefore the separation of  the nuclear  components would require unaccessible   magnetic field
gradients. To overcome this difficulty, experiments have successfully taken advantage of the    called  Optical Stern-Gerlach (OSG) effect  produced by circularly polarized laser beams. The basic idea is that the spin-dependent light shift generated by circularly polarized beams mimics a fictitious magnetic field, which can be used to resolve the
nuclear manifold~\cite{Deutsch1998}. For the ${}^{173}$Yb gas \cite{Taie2010} one OSG beam  was sufficient to separate four of the six nuclear spin  states. The remaining
two nuclear states could be analyzed by reversing the polarization of the OSG beam. For a ${}^{87}$Sr gas, the simultaneous  application of  two OSG beams with opposite circular polarization was required to  resolve all the nuclear spin states~\cite{Stellmer2011}.

An alternative and complementary tool to resolve nuclear spin components uses spectroscopic methods. These  are ideal for AEA thanks  to their narrow intercombination lines. The first demonstration of this technique was achieved using the ultra-narrow ${}^1 S_0$- ${}^3 P_0$ transition in an optical lattice clock \cite{Boyd2006} operated with a cool (at temperature  of a few $\mu$K) but not quantum degenerate  ${}^{87}$Sr gas. The ${}^1 S_0$- ${}^3 P_0$ is only allowed (laser light couples weakly to the clock states) because in the excited state, the hyperfine
interaction leads to a small admixture of the higher-lying P states \cite{Boyd2007}. This small admixture strongly affects the magnetic moment, so that the nuclear $g$ factor of the excited state significantly differs from that of the ground state (i.e $\sim 50\%$ for strontium). The differential $g$ factor was used to resolve all ten nuclear spins in  a bias magnetic field. The spectroscopy was performed  in a deep one dimensional optical lattice designed to operate at the so-called magic wavelength, at which the light shifts on the clock states are equal and the clock frequency is not perturbed by them \cite{Ye2008}. A similar procedure  but using instead the ${}^1 S_0$- ${}^3 P_1$ intercombination line was used
in Ref.~\cite{Stellmer2011} to perform nuclear spin dependent absorption imaging.

A fundamental consequence of the
 SU$(N)$ symmetry  is the conservation of the total number of atoms with
nuclear spin projection  $m_I$, $-I\leq m_I\leq I $. This means that an atom with large $I$ such as ${}^{87}$Sr can reproduce the dynamics of atoms with
lower $I$ if one takes an initial state with no population in  the extra  levels.   Ref.~\cite{Stellmer2011} tested this fundamental  feature by measuring   spin-relaxation  using   the nuclear-spin-state  detection techniques described above. The spin-relaxation rate constant  was found to be   less than $5\times 10^{-15}$ cm${}^3$s$^{-1}$.

\section{Experiments in Optical Lattices:
Realization of a SU$(6)$ Mott Insulator}\label{sec:oplatt}

Optical lattices provide us with a new way of studying ultracold atomic gases by confining them in  periodic arrays that strongly resemble the potential experienced by  electrons in crystaline solids. The optical lattice potential is highly controllable  and can be used to tune the interatomic interactions, the density, the kinetic energy and even the dimensionality of the system by tightly confining the atoms   along specific directions  (see {\it e.g} Refs.~\cite{Bloch2008r,Greiner2008,cazalilla2011} for a review and references therein).

 AEA gases trapped in optical lattices realize the SU$(N)$ generalization of the Hubbard  model~\cite{Cazalilla2009,Taie2010}~\footnote{Henceforth, we
adopt the Einstein convention, which implies that summation must
be implicitly understood over repeated Greek
indices unless otherwise stated.}:
\begin{equation}
H = -t_g \sum_{\langle i, j \rangle}  \left[ c^{\dag}_{\alpha i} c^{\alpha}_j + \mathrm{h.c.}\right] + \frac{U_{gg}}{2} \sum_{i}
n_i(n_i-1), \label{eq:hubbard}
\end{equation} where $\sum_{\langle i,j\rangle}$
stands for summation over nearest-neighbor lattice sites. $c_{\alpha i}$ are fermionic annihilation  operators of $g$ atoms in nuclear spin $\alpha$ at lattice site $i$. The lattice site index $i$
is associated with the vector
$\boldsymbol{R}_i =(R_{x_i}, R_{y_i}, R_{z_i}) $, where $R_{r_i} = m_{r_i} a$ ($r = x,y,z$),
$m_{r_i}$ being   positive integers and $a$ the lattice
parameter.
$n_i = c^{\dag}_{\alpha i}
c^{\alpha}_i$ is the operator that measures
the total fermion occupation (irrespective of the spin)
at the lattice site $i$. The dimensionality
of the lattice, $d$,  and the lattice spacing $a$ are  determined by the number
of counter-propagating laser beams employed to create the lattice potential and the laser
wavelength respectively~\cite{Bloch2008r,Greiner2008}. Equation~\eqref{eq:hubbard} describes the dynamics of a dilute ultracold Fermi
gas hopping between nearest neighbour lattice sites  and interacting only locally. The lattice potential  is assumed  deep enough that
only the lowest Bloch band is occupied by the atoms.  In this regime at most $N$ fermions can occupy the same lattice site.

The Hubbard model  is
written in a form that is manifestly SU$(N)$
invariant. It is characterized by two energy scales,
$t_g$ and $U_{gg}$, which correspond to the kinetic and interaction energy, respectively, and are determined
by the depth of the
optical lattice potential~\cite{Bloch2008r,Taie2010}. $U_{gg}$ is proportional to   $a_s$, i.e.  the
$s$-wave scattering length between two atoms in the ground state. Experimentally, the ratio $U_{gg}/t_g$ can be tuned by varying
the depth of the lattice potential, which in turn is controlled by the  intensity of the laser beams generating the lattice~\cite{Bloch2008r}.

Roughly speaking, at absolute temperatures $T \ll t_g/k_B$, when the kinetic energy dominates (i.e. $t_g\gg U_{gg}$) and  away from special values of the  lattice filling, $n = \langle n_i \rangle$, the system is expected to be a Fermi liquid (see section~\ref{sec:flt}).
On the other hand, when the lattice
filling, $n$, takes integer values $n < N$, and
the interaction energy dominates, i.e. $U_{gg}\gg t_g$,
the hopping of the atoms between lattice sites
is strongly suppressed. This is because, in order to be able to move
around, atoms must pay an energy penalty $\approx U_{gg}$, which at low temperatures $T \ll U_{gg}/k_B$ is not available. Thus, the system becomes a Mott insulator,
for which atom motion is blocked by interactions. This situation is different from the so called band insulator which happens when  $n = N$. In this case the lowest Bloch band is completely filled and  the atom motion is blocked, even in the absence of interactions, by the Pauli exclusion principle.

The experiment reported in Ref.~\cite{Taie2010}, describes
the realization of the SU(6) Hubbard model  by loading an nuclear  spin mixture of $^{173}$Yb atoms in their ground state ($g = ^{1}S_0$)
in a three-dimensional cubic  optical lattice with lattice spacing  $a = 266 \: \mathrm{nm}$.
The lattice was generated by  $d = 3$ mutually orthogonal  pairs of  counter-propagating  laser beams.
In addition to the two terms in Eq.~\eqref{eq:hubbard}, in the experiments there is a confining   potential generated by the Gaussian curvature of the lattice beams. The latter is described by adding to Eq.~\eqref{eq:hubbard} the term:
\begin{equation}
V_{\mathrm{trap}} = \sum_{i}
V_i n_i.
\end{equation}
The trapping potential is well approximated by a harmonic trap, i.e.
$V_i = \frac{1}{2} \sum_{r=x,y,z}  m \omega^2_r a^2/2
\left( \frac{R_{r_i}}{a} \right)^2$, where
 $\omega_r$ is the trap frequency
along the $r = x, y, z$ directions ( for example $\omega \approx 2\pi \times 100$
Hz in Ref.~\cite{Taie2010}).

\begin{figure}
\centering
    \includegraphics[width=110mm]{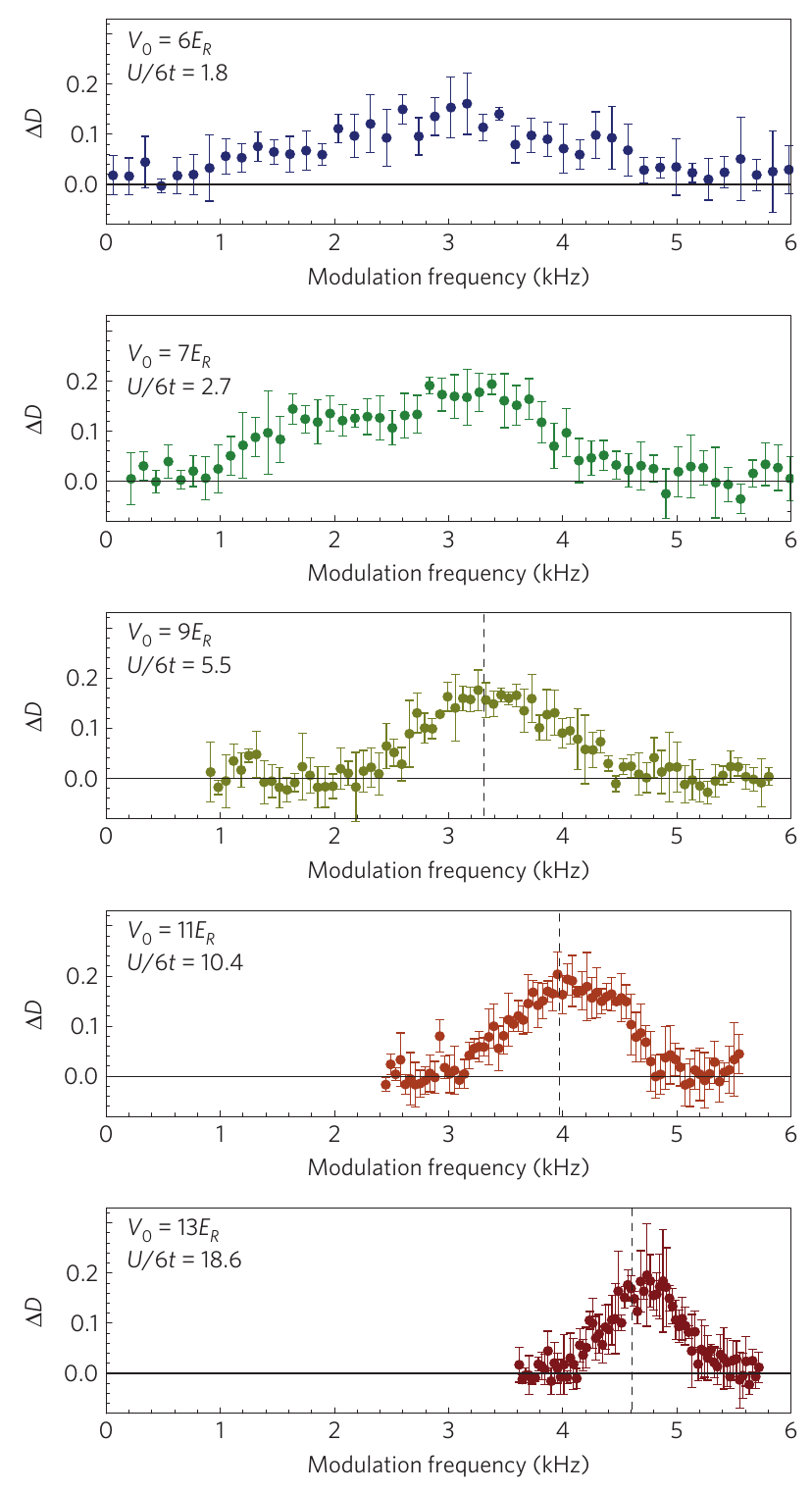}
    \caption{Lattice modulation spectra vs. modulation frequency for increasing
values of the lattice depth (measured in units of the $^{173}$Yb recoil energy, $E_R$). The series shows the emergence of a peak centered around the frequency corresponding to the Mott gap.}
\label{fig:dpr}
\end{figure}

 In order to realize a Mott insulator with SU$(N=6)$ symmetry, the Kyoto group followed the standard adibatic loading procedure used to  create Mott insulators in alkali-metal gases \cite{Bloch2008r}. Specifically, an  ultracold gas
of $^{173}$Yb atoms was first loaded in a 3D dipole trap and then  into a deep  optical lattice by ramping slowly the lattice depth  up to a maximum final value of  $13\, E_R$ ($E_R = \hbar^2 \pi^2/m a^2$
being the recoil energy of the atoms). The loading
was checked to be   adiabatic by reversing
the ramp of the optical lattice and finding that the initial
and final temperatures were  very close to each other.
For the final trapping conditions quoted above
and the initial temperature of the gas ($T_i/T_F \approx 0.2$), the maximum lattice filling was below $2$ atoms per site even at  the center of the trap. This condition is crucial for probing the Mott insulator phase.

 To probe the SU$(6)$ Mott phase and, in particular, to infer its temperature, the Kyoto group  used lattice
modulation spectroscopy~\cite{Stoferle2004,Iucci2006,Iucci2006b,Tokuno2012}. The latter  applies a small periodic (in time) modulation to the  optical lattice depth, which heats  the gas. When the system enters the Mott phase, the injected energy causes  the creation of
holes and doublons, i.e. empty sites and doubly-occupied
sites, respectively.  The doublons production rate (DPR) can be measured by converting the doublons into molecules via  photo-association. The molecules escape very fast from the trap and thus can be detected as  atom loss.  For deep lattices, the
DPR   as a function of frequency exhibits a peaked distribution centered at  the frequency corresponding to
the Mott gap ($\approx U_{gg}/\hbar$ for
$U_{gg}\gg t_g$, see Fig.~\ref{fig:dpr}). Hence,
the lattice modulation provides a direct measurement
of the Mott gap. This technique can be also used
to estimate the temperature of the gas in the lattice,
which  sets the system in the
regime $t^2_g/U_{gg}\ll t_g < k_B T < U_{gg}$~\cite{Taie2010}. Theoretical calculations  based  on  slave particle methods and high-temperature series expansions \cite{Tokuno2012} agreed with the experimental observations.

\begin{figure}
\centering
    \includegraphics[width=140mm]{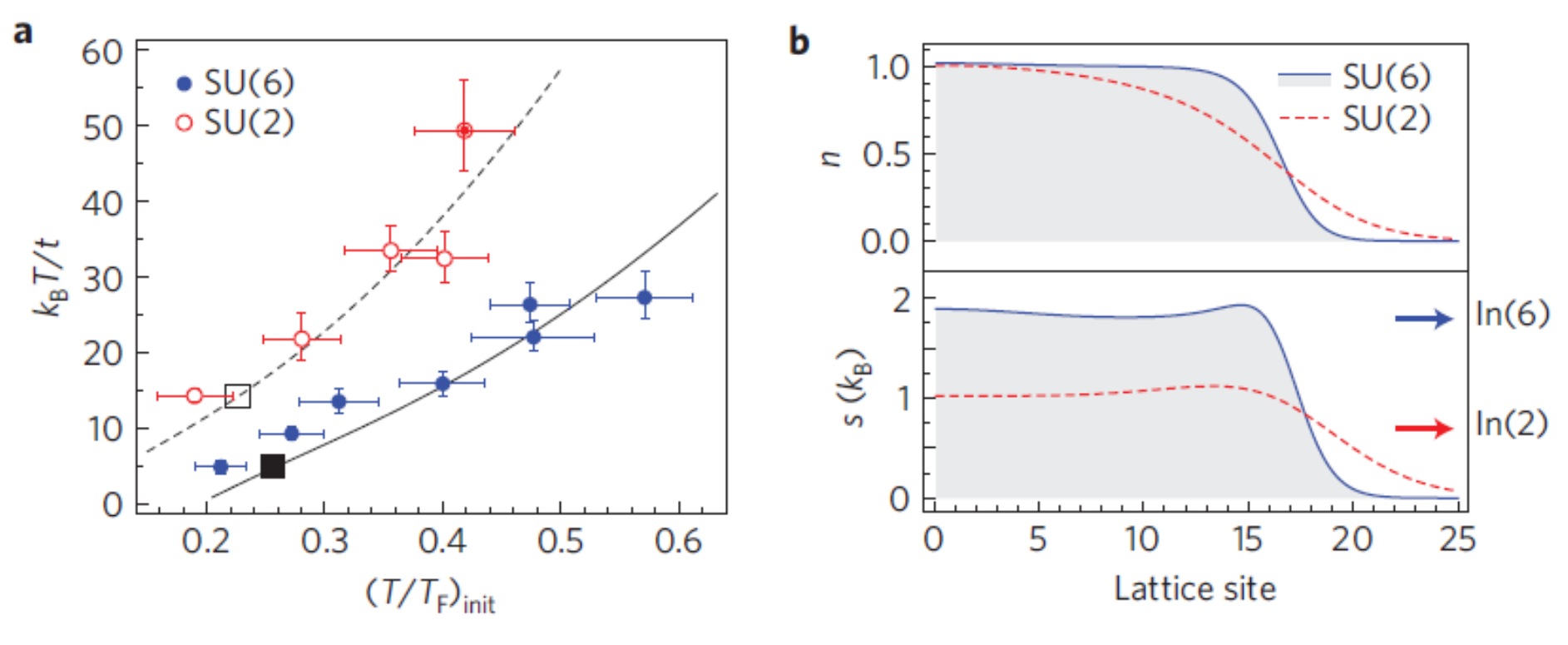}
    \caption{ Adiabatic loading of a SU$(N)$ insulator. From Ref. \cite{Taie2012}: a)Temperatures of SU(6) and SU(2) Fermi gases in the lattice inferred from doublon production rate. The atom number is $1.9(1)\times 10^4$, the lattice depth is 11$E_R$ ($t/h=J_{gg}/h= 63.7$ Hz and $U/h=U_{gg}/h=4.0$ kHz). The dependence on the initial temperature
in the harmonic trap is shown. The solid and dashed lines are the corresponding theoretical curves that assume adiabatic loading into the lattice\cite{Hazzard2012}, and the
square boxes indicate the conditions for which the calculations in b were performed.  b) Calculated density (top) and entropy distribution (bottom) at the lowest measured temperatures for the six and
two-component cases, indicated by squares in a. The maximum spin entropy $\ln(N=6,2)$ is indicated by the arrows.}
\label{Cooling}
\end{figure}

By comparing the  temperature measurements taken for the Mott insulating phases of  SU(6) and SU(2) gases (the latter achieved by optical pumping, remember $N$ can be controlled by initial state preparation) it was found that the final temperature for the SU(6) gas was  a factor of $\sim 2$ or 3 smaller than the one they reached for the SU(2) system (see figure~\ref{Cooling}). The initial $T_i/T_F$ values,  achieved as a result of evaporative cooling, were almost the same for both the SU(2) and SU(6) cases --not the bare temperature--.

 These measurements
were consistent with the theoretical expectations that systems with SU$(N > 2)$ symmetry, adiabatically
loaded on a  lattice,  can be   more efficiently cooled down than SU($2$) systems ~\cite{Cazalilla2009,Hazzard2012}. The cooling can be understood as a direct consequence of the large entropy stored in the spin degrees of freedom in a SU$(N)$ Mott insulator in the  $t_{g} < k_{\rm B}T < U_{gg}$ regime. See Sec. \ref{sec:thoplat} for a more detailed discussion.


\section{Femi liquids and their instabilities}\label{sec:flt}

\subsection{SU$(N)$ Fermi liquid theory and Pomeranchuck Instabilities}

 At temperatures well below the Fermi Temperature $T_F$, in a trap or in an optical lattice of weak to moderate trap depth, a gas of AEA atoms is expected to be a
Fermi liquid. The latter defines a universality class of interacting fermion systems.
As introduced by Landau~\cite{Landau1957} (see {\it e.g} Ref.~\cite{pethick_baym_book} for a review),
Fermi liquids are characterized by the existence of a gapless Fermi surface (FS)
and long-lived low-energy fermionic elementary excitations known as quasi-particles (QP). The QP states can be put in one-to-one correspondence with
the excited states of a non-interacting Fermi gas, which implies that QPs
carry the same quantum numbers as non-interacting particles in a Fermi gas.

  For a uniform SU$(N)$ symmetric AEA Fermi liquid,
the above statements mean that momentum and SU$(N)$ (nuclear) spin
are good quantum numbers, and therefore a QP distribution function
in momentum space,  $n_{\alpha}(\boldsymbol{p})$, can be defined.
At $T = 0$,  the ground state of an unpolarized  three-dimensional gas
of mean density $\rho_0 = N_0/V$ is described by a Fermi distribution QP given by
  $n_{\alpha}(\boldsymbol{p}) = n_0(p) =  \theta(p - p_F)$, where $p_F=  \left( \frac{4\pi^2 \rho_0}{N} \right)^{1/3}$ is the radius of the FS. In order to describe excitations, it is useful to generalize the QP distribution
function to a density matrix, $n^{\alpha}_{\beta}(\boldsymbol{p})$, which
allows us to describe excited states in which the different orientations of the nuclear spin may
be entangled. Following Landau, the grand-canonical free energy (at $T = 0$) of the
excited states can be written as~\cite{Cazalilla2009,ChitovSenechal1995}:
\begin{equation}
F = F_0 + \sum_{\boldsymbol{p}} \left[\epsilon_0(\boldsymbol{p}) - \mu \right]
\delta n^{\alpha}_{\alpha}(\boldsymbol{p})  + \frac{1}{2V} \sum_{\boldsymbol{p},\boldsymbol{p}^{\prime}}  f_{\alpha \gamma}^{\beta\delta}(\boldsymbol{p},\boldsymbol{p}^{\prime})
\delta n^{\alpha}_{\beta}(\boldsymbol{p}) \delta n^{\gamma}_{\delta}(\boldsymbol{p}^{\prime}), \label{eq:landaufnrg}
\end{equation}
where $\epsilon_0(\boldsymbol{p})$ is the bare quasi-particle energy, $\mu$  the
zero-temperature chemical potential, and $\delta   n^{\alpha}_{\beta}(\boldsymbol{p})   = n^{\alpha}_{\beta}(\boldsymbol{p}) - n_0(p) \delta^{\alpha}_{\beta}$ is the deviation of the QP distribution with respect to the ground state. For $p\approx p_F$, $
\epsilon_0(\boldsymbol{p}) = \mu + \frac{p_F}{m^*}(p-p_F)$, where $m^*$ is the QP mass and $\mu$ the zero-temperature chemical potential.

  In Eq.~\eqref{eq:landaufnrg} $f_{\alpha \gamma}^{\beta\delta}(\boldsymbol{p},\boldsymbol{p}^{\prime})$ is the Landau function that describes the (forward scattering) interactions between the QPs. As explained in the Appendix~\ref{app:landau}, the Landau function
can be parametrized in terms of discrete set of Landau parameters $\{F^{\rho}_L, F^{m}_L\}$, where $L = 0, 1, 2, \ldots$ in an integer. The Landau parameters can be obtained, to lowest order in  the gas parameter $p_F a_s$ using the Hartree-Fock approximation. This
yields $F^{\rho}_0 \simeq 2(N-1)p_F a_s/\pi$ and $F^{m}_0 \simeq -2 p_F a_s/\pi$, and vanishing values for  $L > 0$ (hence, $m^* = m$)~\cite{Cazalilla2009}. Recently, they have also obtained to second order in $p_F a_s$~\cite{Yip2013}. The higher order corrections are much enhanced at large N. This means, in particular, that the region at which the Hartree-Fock (HF) results
apply rapidly shrinks with $N$ because the applicability criterion for HF  is $N k_F a < 1$.

For the isotropic Fermi liquid state to be stable, the positivity of the free energy fluctuations to quadratic order requires that
$F^{\rho,m}_{L} > -(2 L + 1)$, otherwise the system
undergoes a Pomeranchuck instability\cite{ChitovSenechal1995,Cazalilla2009, pethick_baym_book}
that can result in a  permanent deformation of the FS, which may or may not
be accompanied by a spontaneous breaking of the SU$(N)$  symmetry. A notable example of Pomeranchuck instability is Stoner instability, which corresponds to the transition from an spin unpolarized (i.e. paramagnetic) to a polarized  gas.
For a system in a trap, where number of particles in each
component is fixed, this transition corresponds to the spatial
segregation of the different nuclear spin components.

Within Landau Fermi liquid theory, the Stoner instability happens if  $F^{m}_0 < -1$. Interestingly, to lowest order in $p_F a_s$,
this criterion is $p_F a_s \simeq \frac{\pi}{2}$, which is the same for all $N$~\cite{Cazalilla2009}. Yet, the analysis based on Fermi liquid
theory of the Stoner instability can be quite
misleading~\cite{Cazalilla2009}, as it predicts a continuous phase
transition for all values of $N$. A more careful treatment
begins by noticing that the order parameter
for $N > 2$ is a traceless hermitian matrix $M$ belonging
to the adjoint representation $N^2-1$ whose components are
$M_{\alpha}^{\beta} \propto  \sum_{\boldsymbol{k}} \langle \left[ c^{\dag}_{\alpha}(\boldsymbol{k}) c^{\beta}(\boldsymbol{k}) - \delta^{\alpha}_{\beta} c^{\dag}_{\alpha}(\boldsymbol{k}) c^{\alpha}(\boldsymbol{k}) \right] \rangle$. Hence,
the change in the Landau free energy at
the Stoner transition can be written as:
\begin{align}
F - F_0 = a_2\:  \mathrm{Tr} \: M^2 + a_3\: \mathrm{Tr}\:  M^3 +  a_4\:  \mathrm{Tr}\:  M^4 + \cdots \label{eq:mfreenrg}
\end{align}
where $a_2 \propto (F^m_0 + 1)$. For $N = 2$, the
second term in the right hand side of ~\eqref{eq:mfreenrg}
vanishes because for a $2\times 2$ traceless matrix $\mathrm{Tr} \: M^3 = 0$. However, this is not so for $N > 2$, which implies
that the Stoner transition is a first-order  (i.e. discontinuous)  transition at the mean-field level~\cite{Cazalilla2009} (for $N = 2$ the Stoner transition becomes discontinuous at low temperatures due to fluctuation effects beyond mean field theory~\cite{Belitz2002,duine2005}).
As a consequence, close to the Stoner transition the system
will exhibit metastability and phase coexistence. Furthermore, this also means that,
at a quantitative level, the Pomeranchuck-Stoner criterion $F^m_0 > -1$ does not provide a reliable estimate of the transition point~\cite{Cazalilla2009}.
Nevertheless, whilst qualitatively correct, the above argument   assumes that the order parameter, i.e. the matrix elements of $M$, remain small in the neighborhood
of the critical point so that keeping the lowest order terms from an expansion in $M$ is enough to capture the transition properties. On the other hand, a direct numerical minimization of the
\emph{total} free energy which does not assume $M$ to be small shows that this is not the case~\cite{Cazalilla_unpub}.  Indeed,  for an ultracold gas with SU$(N > 2)$ symmetry, the Stoner transition appears to be strongly first order, although the conclusion obtained
from the above free energy form remain correct only at the qualitative level.

 Nevertheless, the experimental values of the gas parameter
in a trap (e.g. $p_F a_s\simeq 0.13$ for $^{173}$Yb) are
far from  the critical value corresponding to the Stoner
transition. Furthermore, as explained in section~\ref{sec:optfesh}, the
enhancement of the scattering length by optical means (i.e.
optical Feshbach resonances), breaks SU$(N)$ symmetry
and introduces large atom losses, which may complicate
the applicability of the results discussed above.
Yet, it may still be possible to observe a transition
to a polarized (i.e. spatially segregated) state in
a not too deep optical lattice, as has been
recently suggested by  Monte-Carlo simulations
for two-component mixtures~\cite{MaPingNang2012}.
Or near half-filling in deep optical lattice,
as suggested by a Gutzwiller approximation to
the SU$(3)$ Hubbard model~\cite{Rapp2011}
and  a recent generalization of Nagaoka's
theorem for the SU$(N)$ Hubbard model~\cite{Katsura2013}.

However, although the instabilities discussed
above may not be accessible in the current experimental conditions,
the experimental measurement of the Landau
parameters is still interesting open problem, and
should provide a further confirmation that the SU$(N)$ symmetry
survives many-body effects.  Indeed, the lowest $L$ Landau parameters
can be obtained from the measurement of
the equation of state as it was done recently for the two-component
Fermi gas~\cite{Nascimbene2010,Nascimbene2010b} and from the measurement
of the number fluctuations $\langle (N_{\alpha} - \langle N_{\alpha} \rangle)^2
\rangle$ of the different spin components~\cite{Yip2013}.

\subsection{BCS Instability and Superfluidity}\label{sec:bcs}

 Besides the Pomeranchuck instabilities, the  Fermi liquid state is notoriously unstable against the formation of Cooper pairs, which is known as the BCS (after Bardeen-Cooper-Schrieffer) instability~\cite{Schriefferbook}. Such an instability cannot be described within the framework of Landau Fermi liquid theory because it involves a scattering channel between Landau QPs that is neglected in Landau's theory~\cite{Schriefferbook,ChitovSenechal1995}. However, its importance cannot be
 understated since,
for arbitrarily weak interactions, the Fermi liquid state
is always unstable against Cooper-pair formation~\cite{KohnLuttinger} at sufficiently low temperatures (whether
such temperatures can be experimentally reached is
a separate issue). The angular momentum of the Cooper pairs  as well as the transition temperature depend on the details of the QP interaction, with attractive interactions typically leading to
paring in the $s$-wave channel, and repulsive interactions requiring higher angular momentum channels~\cite{KohnLuttinger}.

 Indeed, multi-component systems exhibit a richer
phase diagram of paired states~\cite{ModawiLeggett1997,Honerkamp2004b,He2006,Cherng2007,Rapp2007,Ozawa2010,Yip2011} than
two component systems~\cite{Giorgini2008,Bloch2008r}.
Below we focus on the case of attractive interactions
and $s$-wave pairing. For repulsive interactions, paring
in channels other than $s$-wave
is also possible at very low-temperatures~\cite{KohnLuttinger,ChitovSenechal1995}, which are
currently beyond the experimental reach. Furthermore,
 $d$-wave paring is also possible  below half-filling
(i.e. when the number of fermions per site
$\lesssim 1/2$) for repulsive  interactions
although  a weak coupling analysis~\cite{Honerkamp2004} shows that the critical temperature rapidly decreases with $N$.

 In order to understand the rich pairing possibilities
of multi-component Fermi gases, let us recall that the $s$-wave order parameter of a paired state  in a uniform gas is
$\Delta^{\alpha\beta} \propto \sum_{\boldsymbol{k}}\langle c^{\alpha} (\boldsymbol{k}) c^{\beta}(-\boldsymbol{k}) \rangle$.
When represented by an $N\times N$ matrix, it corresponds
to an skew-symmetric matrix $\Delta^{T} = -\Delta$,
where $T$ means transposition.
It follows that $\mathrm{det} \: \Delta =
(-)^N \mathrm{det} \: \Delta^T =
(-)^N \mathrm{det}\:  \Delta$. Thus,
for odd  $N$ the determinant vanishes, which implies that there is at least one zero eigenvalue. The corresponding  null eigenvector
$v_{\alpha}$ ($\Delta^{\alpha \beta} v_{\beta} = 0$) determines which component of the mixture remains \emph{unpaired}, i.e. $c_{\mathrm{unp}}(\boldsymbol{k}) = v_{\alpha}c^{\alpha}(\boldsymbol{k})$. This component remains in a Fermi liquid state, wheras the orthogonal
components may be paired or not depending on energetic considerations~\cite{Cherng2007}. In general,  we can rely on Youla's decomposition~\cite{Youla1961,LangAlgebra} and write
$\Delta =  U \tilde{\Delta} U^{T}$, where $U$ is a unitary
matrix and $\tilde{\Delta}$  is a skew-symmetric matrix
for which only the entries $\tilde{\Delta}^{12} = \tilde{\Delta}^{34} \ldots  =
\tilde{\Delta}^{k,k+1}$ with $k \leq N$ are non-zero while
the rest vanishes. Physically,
this means that it is always possible to find a basis
in which component $1$ pairs with $2$,
$3$ pairs with $4$, etcetera,  and  the system  can be described in terms of $k \leq N$ Cooper-pair condensates. Such pairing
states were termed diagonal pairing states by Cherng and
coworkers~\cite{Cherng2007}. For example for $N=3$, two components
pair whereas a third one remains unpaired. In the  weak coupling limit, i.e. for $|p_F a_s| \ll 1$,  the critical temperature has  takes an exponential form similar to the formula obtained in  the two-component mixture case: $T_c = \frac{8\epsilon_F e^{\gamma-2}}{\pi} e^{-\frac{\pi}{2 p_F |a_s|}}$~\cite{ModawiLeggett1997,Honerkamp2004b,Ozawa2010},
where $\epsilon_F = \frac{\hbar^2 p^2_F}{2m}$ is the Fermi energy, and $\gamma \simeq 0.5772$ Euler's constant.
For the entire BEC to BCS crossover, $T_c$ has been
recently obtained by Ozawa and Baym~\cite{Ozawa2010},
following the method of Nozi\'eres and Schmitt-Rink~\cite{NSR1984} to account for the paring fluctuations.
In the BEC limit where $p_F a_s \to 0^+$ they obtained $T_c/T_F \to 0.137$~\cite{Ozawa2010}.

 Nevertheless, we must emphasize that the
existence of $k \leq N$ Cooper pair
condensates does not
rely upon the SU$(N)$ symmetry and entirely  follows from
$\Delta$ being a skew symmetric tensor~\cite{Cherng2007,Yip2011}. On the other hand,  SU$(N)$ symmetry is important and leads to a set of Ward-Takahashi identities that are only satisfied by diagonal pairing  states and not by combinations of them~\cite{Cherng2007}.
Moreover, the SU$(N)$ symmetry plays a crucial role in determining the number and dispersion of the Nambu-Goldstone (NG) collective modes (akin to the Anderson-Bogoliubov mode in the two-component BCS system). This is illustrated using the SU($3$) case in Appendix~\ref{app:ng}, where it is shown that the number of NG modes is not equal to the number of broken-symmetry generators and that  for $N$ odd there are quadratically dispersion NG modes for $N$ odd.

 The existence of quadratic gapless modes and  in particular a gapless unpaired component for $N$ odd
may appear to have important consequences for the superfluidity of the system, according to the Landau criterion~\cite{LandauLifshitz}. This is because the unpaired component and the quadratically dispersing NG modes will cause dissipation when a macroscopic object moves through the fluid.
 However, Modawi
and Leggett~\cite{ModawiLeggett1997} have argued for the irrelevance  of this criterion when applied to such paired
states. As pointed out
by these authors, the superfluid fraction
at zero temperature for these systems remains
finite in spite of the presence of the unpaired
component.

 It is also worth discussing the effects of population
imbalance. Indeed, this is another aspect for which the
behavior of the $N > 2$ systems noticeably deviates
from the $N = 2$ case~\cite{He2006,Cherng2007,Ozawa2010}. The reason can be understood
using the following group-theoretic argument.
As pointed out in the previous section, the magnetization can be represented by a hermitian traceless matrix, $M$. As to
the pairing function, it is a complex rank-$2$ tensor, which can be presented by a matrix $\Delta$. Thus, it is possible to construct a scalar invariant that couples
pairing and magnetization as follows $\mathrm{Tr} \Delta^{\dag} \Delta M = -\Delta^{\alpha\beta} \Delta^*_{\beta\gamma}M^{\gamma}_{\alpha}$ (recall that $\Delta^{\dag}_{\alpha\beta}  =(\Delta^{\beta\alpha})^* = -(\Delta^{\alpha\beta})^*$, see Appendix~\ref{app:sun}). Thus,
the Ginzburg-Landau free energy reads~\cite{Cherng2007,Ozawa2010}:
\begin{equation}
F - F_0 = a_2 \: \mathrm{Tr} \: M^2 + b_2 \: \mathrm{Tr} \: \Delta^{\dag} \Delta  + c_3\: \mathrm{Tr} \: \Delta^{\dag} \Delta M + a_3 \: \mathrm{Tr} \: M^3  \cdots
\end{equation}
where $b_2 \propto (T-T_c)$,   $T_c$ being the critical
temperature of the paired state.
Hence, for $T < T_c$, unless we are dealing with a pairing state such
that $\Delta^{\dag}\Delta \propto \boldsymbol{1}$, the  pairing will lead to a finite
magnetization (i.e. phase segregation in a trapped
system)~\cite{Cherng2007,Ozawa2010}. Note that this is always the case for $N = 2$ as  $\Delta$ is a $2\times 2$ skew-symmetric  matrix, i.e. $\Delta^{\alpha\beta} = \Delta_0\: \epsilon^{\alpha\beta}$ ($\epsilon^{12}=-\epsilon^{21}=1$) and therefore $\Delta^{\dag} \Delta = |\Delta_0|^2 \: 1$.
However, this condition is not generally met for $N > 2$ and  in particular, never when $N$ is odd. The additional term in the free-energy coupling paring and magnetization is also responsible
for turning the transitions between different diagonal paired  states into first-order transitions~\cite{Cherng2007}. Generic phase diagrams for $N=3,4$
have been obtained by Cherng, Refael, and Demler
in Ref.~\cite{Cherng2007}. For $N = 3$, the phase
diagram in entire BEC to BCS crossover has been
computed by Ozawa and Baym~\cite{Ozawa2010}.

 Finally, we should mention that attractive interactions
in systems with $N > 2$ can yield phases involving more complicated bound states like trions, which would correspond to the Baryons of QCD. This possibility has been studied  for three-species gases loaded in an optical lattice~\cite{Rapp2007,Klingschat2010}.

 In closing, we remark that the observation of the paired and trionic phases described above relies on the possibility of controlling the sign of the atomic interactions. Whereas this certainly is possible for both  alkali atoms, using magnetic Feshbach resonances~\cite{Giorgini2008,Bloch2008r}, and AEA, using optical resonances (see section~\ref{sec:optfesh}), both methods break the emergent unitary symmetry of the gas. Thus, how much of what has been described in this section remains valid  depends on the magnitude  of interaction
anisotropies, which set the temperature scale
above which the SU$(N)$ symmetry remains a good approximation~\cite{Cherng2007}.

\section{Alkaline-Earth Atoms in Optical lattices}\label{sec:thoplat}

\subsection{Cooling on the lattice}


Although Mott insulating behavior has been experimentally demonstrated \cite{Taie2010} (see Sec.~\ref{sec:oplatt}), it is of great importance for the quantum simulation program to be able to cool down the optical lattice system to a regime  where $k_{\rm B}T <t^2_{gg}/U_{gg}$. This necessary for observing effects of magnetic exchange. Currently the achieved temperature in
experiments is still in the range $t^2_{gg} /U_{gg} < k_{\rm B}T <U_{gg}$. Although, this is similar to the issues
encountered when studying the SU(2) Hubbard model with cold
alkali gases, recent investigations suggest that the large spin degeneracy present in SU($N$) systems can help to reach colder temperatures in fermionic AEA. In particular,
Ref.\cite{Hazzard2012} studied the finite-temperature Mott-insulator to Fermi gas crossover, in the regime where $k_{\rm B}T>t_g$ by performing a high-temperature series expansion, together with a local density approximation assumption to deal with the external harmonic potential.  It was thus shown that  the final temperatures, achievable by the standard
experimental protocol of adiabatically ramping up the lattice from a weakly interacting gas in a trap, can yield substantially colder Mott insulators. For example, for fixed particle numbers and fixed initial temperatures, relevant to current  experiments, it was shown that increasing $N$  from 2 to 10 can  lead  to Mott insulators  more than a factor of five colder. Furthermore, if the initial entropy, instead of  the temperature, is what is held fixed, the adiabatic procedure  can lead to even better cooling  for all $N$.   The latter case seems to be experimentally relevant  because the Pauli blocking
effect on evaporative cooling depends on entropy, $S_i\propto T_i/T_F$, with $T_F$ the Fermi temperature, rather than the
bare temperature  \cite{Taie2010} .

The cooling can be understood as a direct consequence of the rapidly
increasing  entropy in a SU$(N)$ Mott insulator in the  $t_g < k_{\rm B}T < U_{gg}$ regime. For the  $n = 1$ case, the entropy per particle grows as $S_f\propto \ln{N}$, since
each of the $N$ flavors is equally likely to occupy a site. For
the experimentally relevant range of $N\leq 10$, the logarithm grows faster
than the  entropy of a quantum degenerate non-interacting gas in a 3D trap which scales for fixed initial temperature as, $S_i\propto N^{1/3}$.

The possibility of reaching colder temperatures in the regime $t_g < k_{\rm B}T < U_{gg}$ by storing entropy in the nuclear spin  degrees of freedom  is encouraging. However, the real motivation is the exploration of exotic SU$(N)$ magnetism, which requires temperatures colder than $t_g^2/U_{gg} < t_g$ for $U_{gg} \gg t_g$. Whether or not large $N$ can help us  to reach this regime is a crucial, but at the same time challenging question. Recently, in Refs. \cite{Bonnes2012,Messio2012},   Quantum Monte-Carlo  methods  supported by  analytic \cite{Lee1994} and DMRG (Density matrix renormalization group methods) calculations \cite{Manmana2011}, showed that
 after adiabatically loading a weakly interacting gas into an array of one-dimensional chains, the final temperature decreases with increasing $N$ even in the regime   $k_{\rm B}T<t_g^2/U_{gg}$. According to those calculations, for current initial  conditions, such  adiabatic loading procedure can allow us to
reach  temperature scales at which interesting magnetic physics happens, for example the onset of Luttinger
liquid behavior and  ground-state  algebraic magnetic correlations \cite{Assaraf1999,Manmana2011}.
The cooling is a consequence of the rapid growth of the entropy with $N$,  in  the one-dimensional  SU$(N)$ Heisenberg model (See Sec.\ref{heis}). At low $T$ the entropy  scales as $S_f \propto N(N-1)$ \cite{Lee1994}, even faster than its  corresponding  entropy in the high$-T$ limit, where it scales as  $S_f \propto \ln{N}$, as discussed above. The quadratic grow  can be derived  from
the  exact solution  \cite{Sutherland1975} and the fact that there are  $N-1$ branches
of elementary excitations  all with  the same velocity
$v \propto 2\pi/N$ at small momentum.  The quadratic growth of $S_f$ with $N$ brings the temperature of the system down with increasing   $N$  and into the  region where ground-state-like correlations start to  develop. This was shown in Ref.~\cite{Bonnes2012} by computing the relevant spin-spin correlations at finite $T$ and comparing them to the ones expected for the ground-state from DMRG calculations \cite{Manmana2011}. Refs. \cite{Bonnes2012,Messio2012} also showed that starting from currently achievable temperatures, after adiabatic loading  the gas, signatures of short range  magnetic ordering could be seeing in the spin structure factor for $N\geq 4$. These calculations suggest that it should be
possible to explore features of SU$(N)$ quantum magnetism already
in ongoing experiments with AEA.

\subsection{The SU$(N)$ Hubbard model at weak to intermediate coupling}

\begin{figure}{}
\centering
    \includegraphics[width=140mm]{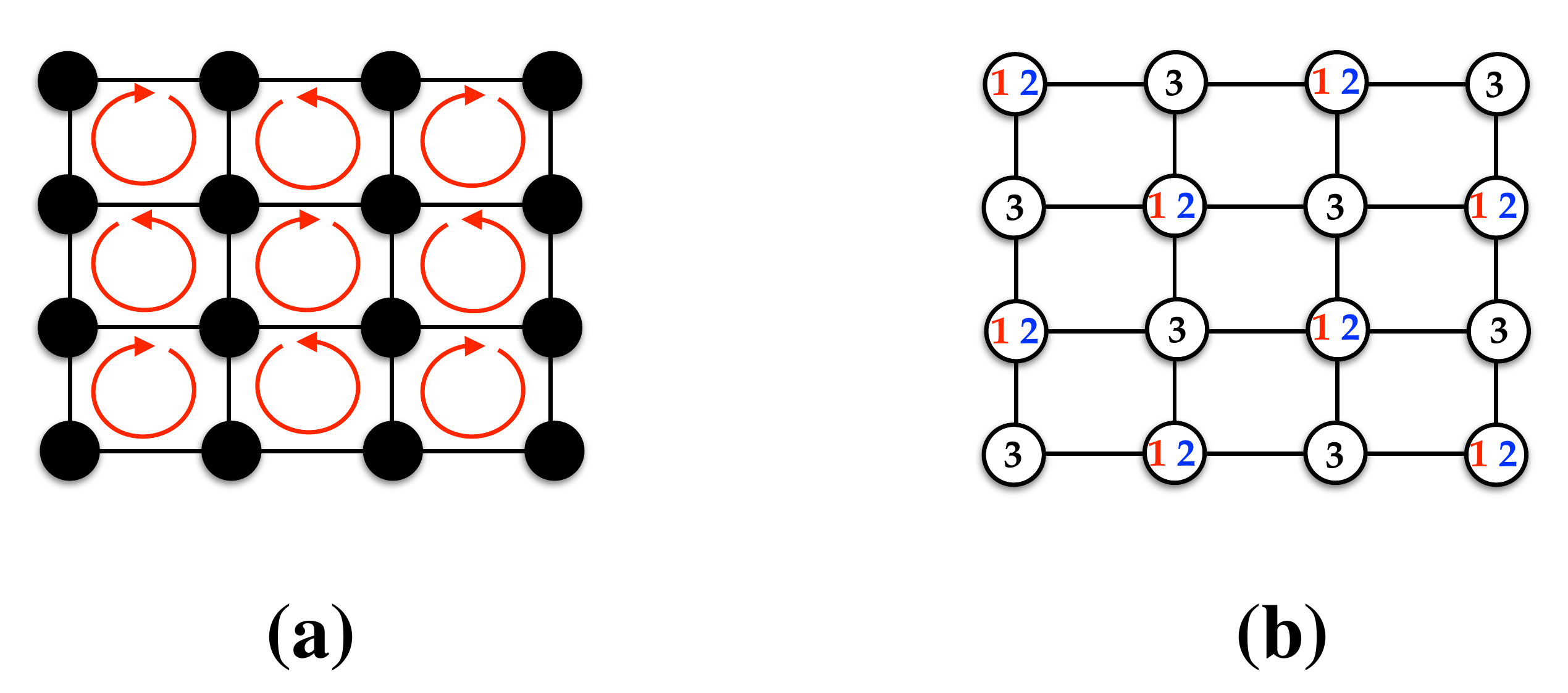}
\caption{Cartoon of  (a) the staggered flux (SF)  and (b) the flavor density-wave (FDW) phases of the SU$(3)$ Hubbard model.  The FDW phase can be regarded as a generalization of the Ne\'el order for $N > 2$.
The SF phase does not break SU$(N)$ symmetry but breaks time-reversal invariance.}
\label{fig:ssfvsfdw}
\end{figure}

 The SU$(N > 3)$ Hubbard model is expected to exhibit a phase diagram  richer than its SU$(2)$
symmetric counter-part. In the weak to intermediate coupling limit,  this  phase diagram has been expored
by Honerkamp and Hofstetter~\cite{Honerkamp2004} for both the attractive and repulsive cases, and by the same authors~\cite{Honerkamp2004b} as well as Rapp \emph{et al.}~\cite{Rapp2007} for $U < 0$. Since work on the attractive case has been already reviewed in section~\ref{sec:bcs}, in this section we focus on the repulsive Hubbard model ($U_{gg} > 0$).

 Besides the Fermi liquid phase that should be stable for $U_{gg}/t_g \lesssim 1$ and  lattice fillings well away from incommensurability, the  SU$(N)$ Hubbard model with repulsive interactions
can display various types of ordered phases. Some of those phases  break the lattice translation symmetry and may or may not break the SU$(N)$ symmetry at the same time. In this respect, they are different from the the phases discussed in section~\ref{sec:flt}, whose order parameters have no spatial dependence (for a uniform system) because these phases do not spontaneously break  translational invariance.

Perhaps the most spectacular example of the above type of phenomena is a phase that breaks lattice translation symmetry without breaking SU$(N)$,  known as the staggered flux (SF) phase ( Fig. \ref{fig:ssfvsfdw}a). This phase was obtained as a mean-field solution of the Hubbard model shortly after the latter gained relevance as the minimal model for the high-$T_c$ cuprate superconductors~\cite{Affleck1988,Marston1989}. It has been postulated as candidate to explain the anomalous pseudo-gap phase of these materials~\cite{Ivanov2000}. The mean-field solution obtained by Marston and Affleck~\cite{Affleck1988,Marston1989} is the exact ground state of the SU$(N)$ Hubbard model for a 2D half-filled lattice (filling fraction $n = \langle n_i \rangle = N/2$) in the  $N\to +\infty$ limit~\cite{Affleck1988,Marston1989}. Therefore,
it is expected~\cite{Honerkamp2004,Cazalilla2009} that it can be realized using ultracold gases as values of $N$ can be as large as $10$  using $^{87}$Sr (Table~\ref{table1}). However, as pointed out by Honerkamp and Hofstetter~\cite{Honerkamp2004}, at values of $N \lesssim 6$, a functional renormalization-group analysis (see also~\cite{Rapp2011}, for a recent variational study) shows that another phase, known as a flavor density wave (FDW) phase (Fig.~\ref{fig:ssfvsfdw} a) is favored over the SF phase. Like the SF phase, the FDW phase also breaks lattice translational symmetry. However, unlike the SF phase, it also breaks SU$(N)$ symmetry. According
to Ref.~\cite{Honerkamp2004}, $^{173}$Yb is  on the
borderline for the stabilization of the SF phase, whereas
$^{87}$Sr is probably a better candidate. It also worth mentioning that the SF phase is characterized by an Ising order parameter which is the direction of the angular momentum associated with the fermion current in each plaquette of the 2D square lattice. Thus, in 2D, the long-range SF order is stable at finite temperatures,  which may facilitate its observation using ultracold atoms.

  However, one major challenge for the observation of these phases is, not only their relatively low ordering temperatures (compared to $t_g$), but the requirement of a lattice fillings at or near the half-filled lattice ({\it i.e. }$n = \langle n_i \rangle \simeq N/2$). For large $N$ this requires a relatively tightly
confinement trapping potential so that large $n$ plateaux can form at the center of the trap~\cite{Cazalilla2009}. However, under such circumstances, it is not clear how stable such the lattice system would be against inelastic losses. For instance, using $^{173}$Yb, a half-filled insulating plateau containing $N/2 = 3$ atoms per site can be reached at the center of the trap~\cite{Cazalilla2009}. However, the existence of such plateau makes the system very susceptible to three-body recombination and the unwanted heating effects associated with it. A precise experimental determination of the lifetime of a high-filling optical lattice for common AEA is in order. Furthermore, on the theory side, not much is known about how such phases, and in particular the SF phase can be detected in the optical lattice setup.

\begin{figure}[b]
\centering
    \includegraphics[width=110mm]{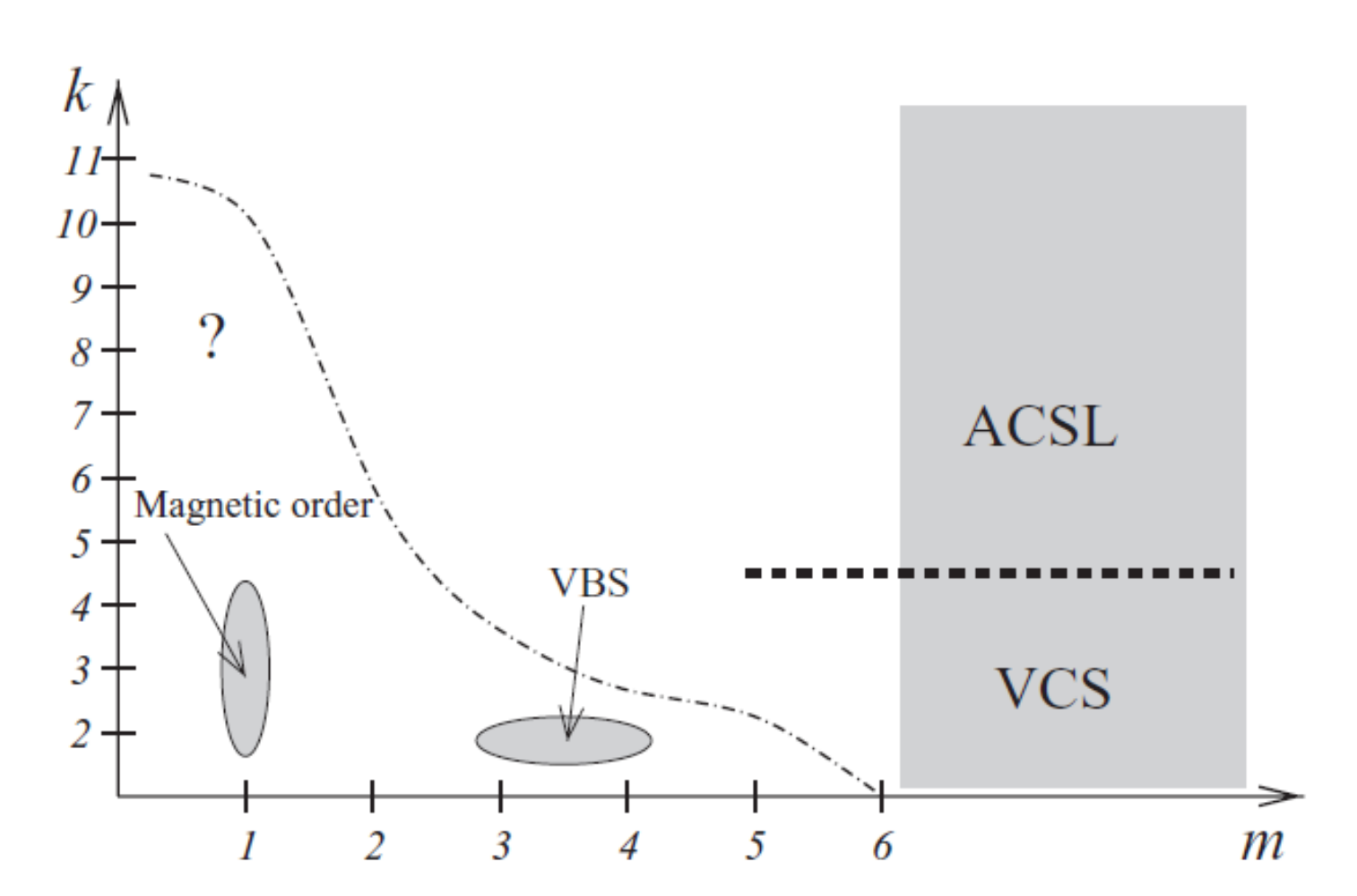}
    \caption{ Phase diagram of the SU(N) Heisenberg model in two
dimensions on the square lattice with $N = m k$, taken from Ref.\cite{Hermele2011}.
$m$ is the number of fermions per site, and $k$ is the number of sites needed  to form a singlet. Regions where there
is substantial evidence for a given ground state, or where the ground
state is known, are shaded. The Abelian chiral spin liquid (ACSL)
and valence cluster state (VCS) regions on the right are established
by large-$N$ analysis; the boundary between these regions in large
$N$ is shown by a dashed line. For $k = 2,m=1$,  the Neel state is the
well-known ground state. There is also evidence for magnetic order
at $k = 3,m=1$ \cite{Toth2010} and $k = 4, m=1$ \cite{Corboz2011}. Valence-bond
solid (VBS) order was found for $k = 2$
and $m = 3,4$  \cite{Assaad2005}. The dashed-dotted line separates the range
of parameters beyond the reach of current experiments (above and to
the right of the line) and the range within the reach of the experiments
(below and to the left of the line). The experimentally relevant part
of the phase diagram with the greatest potential for novel ground
states, in particular, the Abelian chiral spin liquid, is indicated with a
question mark. }
\label{diagram}
\end{figure}
\subsection{Strong coupling limit: The SU$(N)$ Heisenberg model}\label{heis}

As we have discussed already in section~\ref{sec:oplatt},
when AEA  in their ground electronic  state are loaded into a deep optical lattice they provide us with an accurate realization of the SU$(N)$ Hubbard model (see Eq.~\ref{eq:hubbard}).
In the limit of large $U_{gg}/t_{g}$ and for integer filling fractions the system becomes a Mott insulator. In this regime  the motion of the particles only takes place virtually, since adding or removing a particle  at a giving lattice site costs energy, and the Hamiltonian reduces  to an
effective spin Hamiltonian. Assuming a translational invariant system for simplicity (setting $V=0$), the effective model obtained by second order degenerate perturbation theory is the SU$(N)$ Heisenberg model \cite{Gorshkov2010}
\begin{equation}
H= \frac{2 t^2_g}{U_{gg}} \sum_{\langle i,j\rangle } S_\alpha^\beta(i)  S_\beta^\alpha(j),
\end{equation}
where the spin operators $S_\alpha^\beta(i) = c^\dagger_{\alpha,i} c^{\beta}_{i}$ which satisfy the SU$(N)$ algebra $[S_\alpha^\beta(i),S_\gamma^\delta(j)] = \delta_{ij} (\delta_{\beta \gamma} S_\alpha^\delta - \delta_{\alpha \delta} S_\beta^\gamma)$ (see Appendix~\ref{app:sun}).

 Like the Hubbard model reviewed above, the SU$(N)$ Heisenberg model can also display a rich phase diagram. The phases depend on $N$, the filling fraction $n = \langle n_i \rangle$, dimensionality and lattice geometry. The parameter $k\equiv N/n$,  chosen to be an integer greater than unity, plays
a key role in the analysis of the phase diagram: $k$ is the minimum
number of sites needed to form a SU$(N)$ singlet.  The  one dimensional chain with $n=1$  admits an exact solution for all $N$ \cite{Sutherland1975} and its phase diagram is well understood. Nevertheless, the phase diagram of the 2D model is complex and remains unknown to a great extent. The phase diagram  predicted in Refs.~\cite{Hermele2009,Hermele2011} for a square lattice is shown  in Fig.~\ref{diagram}. There  $m$ labels the filling fraction (i.e $m=n$). The dashed-dotted line separates the range
of parameters beyond the reach of current experiments (above and to
the right of the line) and the range within the reach of the experiments
(below and to the left of the line covering the region $N\leq 10$  and $n\le 5$ ). The predictions for the quantum phases, based on a large-$N$ expansion  and thus valid in the limit  $N \gg 1 $  for $k$ finite, have been shaded, as well as regions  where the ground state is known.

 The known regions correspond to   the well established $N=2$  and $n=1$  or $k=2$ case, where anti-ferromagnetic  long range order is favored, and  the ground state is the so called  Neel state. The generic case $k=2$  shares with SU$(2)$  the crucial property that two adjacent spins can form
an SU$(N)$ singlet, and has  been studied extensively as a
large-$N$ generalization of SU$(2)$ magnetism \cite{Read1989,Affleck1988,Marston1989,Rokhsar1990}. Those studies
found that  under very general conditions in the large-$N$ limit, the ground state is non-magnetic and  spontaneously breaks lattice symmetries. It is formed by  two-site singlets and referred to as a  valence-bond solid (VBS). Quantum Monte Carlo simulations done for $N=3,4$  have  confirmed that the  ground state remains a VBS  even at finite $N$.   Numerical studies of the cases  $N=3,4$ but  $n=1$  (or $k=3,4$) in a square lattice \cite{Toth2010,Corboz2011} on the contrary  provide  strong evidence of  magnetically ordered ground states.

 The large-$N$ expansion  predicts  two different ground states  depending on $k$:  a valence cluster state (VCS) formed  by tiling the lattice with multisite singlet clusters  for $k<5$  and an Abelian  chiral spin liquid (ACSL), which is  a spin-system analog of a Fractional  Quantum Hall state \cite{Kalmeyer1987,Kalmeyer1989, Wen1989}, for $k\geq5$. A VCS is non-magnetic and breaks lattice symmetries. The ACSL  spontaneously breaks parity and time-reversal symmetry,  supports excitations with fractional quantum numbers and
statistics, and has gapless chiral edge states that carry
spin.

Although  we have focused our analysis of the phase diagram  of the  SU$(N)$-Heisenberg  model on a  square lattice,  which is the simplest to generate in experiments, it is important to mention that it is expected to be even richer  in more generic lattice geometries. For example numerical investigations  of the SU$(3)$-Heisenberg  model in a triangular lattice predict a perfectly ordered three-sublattice  state  \cite{Launchli2003}.
On a honeycomb lattice, the $SU(3)$ case has been shown to have a dimerized,magnetically ordered state
 \cite{Zhao1212,Lee2012,Corboz2013h}, and  it has been also predicted that the SU(6) case   becomes
 a  ACSL using a large $1/N$ expansion \cite{Szirmai2011,Corbozh2012}. Whether or not the ACSL remains the ground state in the experimentally relevant part
of the phase diagram, $k=N$ and $n=1$ is not unknown and needs to be validated by experiments.



\section{Other exotica: Physics beyond the SU(N) Heisenberg model}\label{sec:other}

In this section we present an overview of some of the recently  proposed exciting  physics that  near term AEA experiments could  explore. Most of those proposals take advantage of the long life time of the $^3 P_0$ state, in addition to the  SU$(N)$ symmetry in the nuclear spin levels.

\subsection{Orbital magnetism} The  possibility to independently manipulate the ${}^1S_0$ and $^3 P_0$ states  by laser light and therefore to construct identical or different optical lattices for the two states \cite{Daley2008} allows for  the  simulation of two-orbital-SU$(N)$ Hamiltonians which rely on the interplay of charge, spin and orbital degrees of freedom. The electronic clock states  play the role of the orbital degree of freedom and the corresponding nuclear spins provide the spin degree of freedom. The investigation of orbital physics using alkali-metal atoms has of course also been considered. For example,
a natural  way to add orbital degrees of freedom is to encode the spin  in their internal hyperfine degrees of
freedom, and the orbital degree of freedom in different lattice bands. However, one important limitation of this approach is that  the occupation of
 excited lattice bands is, at best, metastable \cite{Bloch2008r}. Spin mixtures in alkali atoms in which the orbital degree of freedom is encoded in different type of atoms has been thought as
an alternative to explore orbital physics without the necessity of populating higher bands. In this
case one can easily impose an optical lattice which acts differently on the two species of atoms owing to their different optical properties. However, in this case atom distinguishability only gives rise to pure density-density interactions without direct spin interconversion. The   emulation of orbital physics by using the lowest band orbitals   of independent optical lattice felt by the  ${}^1S_0$ and $^3 P_0$  states  does not have any of the metastability issues of higher bands, and allows also for nuclear spin flip processes.
Collisional relaxation of the electronic excite states must however  be considered \cite{Ludlow2011,Bishof2011}. A possible way to deal with it is to work in the regime where there is only one ${}^3P_0$ atom per lattice site.

The implementation of the two-band SU(N) Hubbard model with alkaline earth atoms opens untapped opportunities \cite{Gorshkov2010,Foss-Feig2010,Foss-Feig2010b},  including the implementation of a SU(N) generalization of the SU(2)-Kondo lattice model (one of the canonical models used to study strongly correlated electron systems, such as manganese oxide perovskites \cite{Tokurabook} and rare-earth and actinide compounds classed as heavy-fermion materials \cite{Coleman2007}) and a SU(N) generalization of the N=2 Kugel--Khomskii Hamiltonian (used to model the spinorbital interactions  in transition-metal oxides with the perovskite structure \cite{Tokura2000}). Just recently it was also pointed out that a SU(N)-Mott insulator with one
ground state atoms and one excited state atoms on each site of  a square lattice is likely to realize  a non-Abelian Chiral spin liquid with a quantum statistics  sufficient for universal
quantum computations \cite{Hermele2011,Freedman2004}. Note that other non-Abelian states such as the fractional quantum Hall
state at the filling fraction 5/2 \cite{Moore1991,Willett2010} or a   variety of setups involving Majorana fermions \cite{Alicea2011} are   not rich enough to support universal quantum computation \cite{Nayak2008}. Recent numerical studies of   the phase diagram of the SU$(4)$ Kugel--Khomskii model
in a honeycomb lattice predict   a quantum spin-orbital liquid ground state\cite{Corbozh2012}.

\subsection{Artificial Gauge fields}

 Atoms are neutral particles and thus they do not experience Lorentz forces in the presence of electromagnetic fields. Recently, it has been demonstrated that  when a neutral atom is illuminated with  properly designed laser fields,
its center-of-mass motion can mimic the dynamics of a charged particle. This is the basis of the so called  artificial (synthetic) Gauge fields for atoms \cite{Dalibard2010}. Although there has been important advances in implementing those ideas in alkali atoms by coupling their internal or motional states  with laser light \cite{Spielman2011,Lin2011,Cheuk2012,Kennedy2013,Miyake2013,Aidelsburger2013} spontaneous emission of the excited levels always imposes limitations. AEA have been pointed out to be ideal for synthetic gauge field implementation \cite{Gerbier2010} thanks to the long lifetime of the excited state, its reduced spontaneous emission rate and   the possibility
of generating  anti-magic lattice potentials for the clock states --the clock states feel exactly the same lattice but with an opposite sign--. The latter has been shown also to facilitate the implementation the so-called optical flux lattices \cite{cooper2011,Cooper2013}. In addition the large number of decoupled nuclear spin degrees of freedom  could  facilitate the implementation of SU(N) non-Abelian gauge fields and spin-orbit Hamiltonians exhibiting rich quantum dynamics  and connections to high energy physics \cite{Banerjee2013,Goldman2013}. For the use of AEA for artificial gauge field implementation however, collisional relaxation of the electronic excite states could impose important limitations and  must   be considered \cite{Ludlow2011,Bishof2011}.

\section{Atom-light hybrid systems and many-body physics in optical  clocks}

 Recent advances in modern precision laser spectroscopy,
with record levels of stability and residual laser drift less than  mHz/s \cite{Nicholson2012,Matt2011,Bloom2013,Ludlow2013} are crucial developments that  are allowing us to deal with
AEA clocks operated at a very different  conditions than those ones dealt with just few years ago.
 The level of energy resolution achievable in current atomic clocks is now providing  the required capability to systematically
spectrally resolve  the complex excitation spectrum of an interacting many-body  system. This was certainly  not the case in prior
clock experiments where interaction effects were subdominant. Optical atomic clocks operating with AEA are thus becoming a new laboratory for the exploration of non-equilibrium many-body phenomena with capabilities not foreseen before\cite{Rey2009,Gibble2009,Lemke2011,Bishof2011,Swallows2011,FossFeig2012,Martin2013,Rey2014}.

Moreover the combination of this new regime of ultrastable atomic
dipoles with optical cavities, is predicted to become a  pathway for realization
of exotic states of matter and light. The idea here is to make the leap to
using light to mediate interactions between atoms,
impose coherence, and/or directly drive dynamics
through strong coupling to matter \cite{Meiser2009,Olmos2013}. The long-lived dipoles will allow coherent
interaction of many atoms with a single optical cavity
mode over an extraordinarily long time, generating
strong correlations. Experiments performed using Raman transitions in alkali vapors to mimic the ultrastable  alkaline-earth dipoles, which have observed superradiant behavior maintained with less than one photon in the cavity, provide first principle demonstration of this outstanding capability\cite{Bohnet2012}.

\section{Summary and Outlook}

 Much has been achieved since the first  time alkaline-earth atom gases were brought to quantum degeneracy. The creation of Bose-Einstein condensates  rapidly led to the production of quantum degenerate Fermi gases. The latter, as we have discussed above, exhibit an emergent SU$(N)$ symmetry, which makes of these gases rather unique many-body systems. Since this fact was pointed out, the field has evolved rapidly leading to the  creation of a SU$(6)$ Mott insulator~\cite{Taie2010} and, very recently, to the realization  of arrays of  one-dimensional   and quantum degenerate ultracold $^{173}$ Yb  gases~\cite{Fallani2014}. These experiments
have also  demonstrated that, thanks to the large entropy that can be
stored in the nuclear spin degree of freedom of the AEA gases,
there is much room for improvement
in the quest for cooling down AEA to lower and lower temperatures (entropies)
using conventional methods such as
sympathetic cooling and adiabatically loading into the lattice.

We point out nevertheless that, regardless all the great progress,  what has been experimentally
 demonstrated so far~\cite{Taie2010,Fallani2014} is just fairly interesting physics related to the ``charge'' degrees
 of freedom. The real challenge associated with  the observation of
quantum magnetism and many of the other exotic phases that
have been described in previous sections  still needs to be overcome.
Those phases should become stable to thermal fluctuations
well below the hopping temperature scale  $\sim t_{g}/k_B$, and typically at $ k_B T   \ll  t^2_{g}/U_{gg}$ for the Hubbard model of section~\ref{sec:oplatt}. As we have emphasized above, we expect that the large spectral degeneracies introduced by the enlarged
SU$(N)$ symmetry will bring about new phenomena which have no counterpart in  the two-component systems. Some hints
of these differences have already shown up in the experiments~\cite{Taie2010,Fallani2014}, but there is more to come if we can
find a way to remove the entropy from the nuclear degree of freedom.
This is a challenge that will require new ideas, perhaps different from those applicable ultracold gases of alkali atoms.

As we have seen, turning other parameters like the interactions
in AEA also requires using different methods like optical Feshbach resonances. Unfortunately, the latter generally spoils the emergent SU$(N)$ symmetry that makes these gases so special. New ideas are also required in this regard. And if an  efficient and versatile way is found to tune the interactions while respecting the SU$(N)$ symmetry, this will open the door to the exploration of superfluidity and ferromagnetism in these systems.
 The landscape associated with  phases, as we have described in section\ref{sec:flt} will be rather rich, exhibiting interesting excitations and topological defects as well as discontinuous phase transitions between them. Those can lead to spectacular phase segregation patterns (i.e. domains) in a trap. On a different but complementary direction, although the potential use of  the exquisite  precision of optical lattice clocks to probe AEA manybody physics has started, there is still lots of room for improvement.

 But in spite of the limitations of the present, it is important to emphasize  that seeds for a bright future of the field have been already planted. We strongly believe that there is much more to come, and hopefully many serendipitous  discoveries are waiting for us.  Some of such discoveries may come in the form of new phases of matter, which do not fit into the relatively narrow framework that we have outlined in this article. Or they may come by exploring non-equilibrium phenomena with AEA gases. Indeed, this is a field that, compared to what has been already achieved using alkali gases, remains largely unexplored at the time of writing this article. And as it happens for equilibrium phenomena, we have a new parameter to play with, namely $N$ (or $1/N$, depending on the point of view). In conclusion, we hope that this review will become the starting point for many of the bold minds wanting to explore these fascinating new systems.

\section{Acknowlegments}

 We thank Karlo Penc, Sungkit Yip, and Yoshihiro Takahashi for useful comments on the manuscript  and discussions. MAC acknowledges support from NSC and NTHU (Taiwan). AMR acknowledges funding from NIST, JILA-NSF-PFC-1125844, NSF-PIF, ARO, ARO-DARPA-OLE, and AFOSR.

\appendix

\section{Brief Digest of $\mathrm{SU}(N)$ Group Theory}\label{app:sun}

 In this Appendix we briefly review the most important facts about the special unitary group  $\mathrm{SU}(N)$.
 We begin by defining it. To this end, let us first introduce a $N$-dimensional linear space of complex vectors denoted as
 $\boldsymbol{\psi}^T = (\psi^1, \ldots, \psi^N)$, where $T$ means transpose, and the components   $\psi^{\alpha}$ ($\alpha = 1, \ldots, N$) are complex numbers. In this linear space, we  define the scalar product between two vectors $ \boldsymbol{\psi}$ and $\boldsymbol{\chi}$ as $\langle \boldsymbol{\psi} | \boldsymbol{\chi} \rangle =  \sum_{\alpha}(\psi^{\alpha})^* \chi^{\alpha}$. In order to lighten the notation, we introduce the dual of the vector $\boldsymbol{\psi}$
defined by $\psi_{\alpha} = (\psi^{\alpha})^*$. This allows to
write  $\langle \boldsymbol{\psi} | \boldsymbol{\chi} \rangle = \psi_{\alpha}\chi^{\alpha}$, where  repeated upper and lower indices are to be summed over, unless stated otherwise. Finally, the norm of  $|\boldsymbol{\psi} \rangle$ can be defined as $\sqrt{\langle \boldsymbol{\psi}|\boldsymbol{\psi}\rangle} = \sqrt{\psi_{\alpha}\psi^{\alpha}}$. Let next us consider the linear transformation
\begin{equation}
\tilde{\psi}^{\alpha}=  U^{\alpha}_{\beta} \psi^{\beta}. \label{eq:sun}
\end{equation}
Hence, the dual $\tilde{\psi}_{\alpha} = (\tilde{\psi}^{\alpha})^* =
(\psi^{\beta})^* (U^{\beta}_{\alpha})^* = \psi_{\beta} (U^{\dag})^{\beta}_{\alpha}$,
where have employed that $(U^{\dag})^{\alpha}_{\beta} = (U^{\beta}_{\alpha})^*$,
where $U^{\dag}$ is the hermitian conjugate of the matrix $U$.

 We are now ready to define the $\mathrm{SU}(N)$ group as the set of  \emph{linear} transformations  that preserve the  norm of vectors.  Mathematically, $\langle \boldsymbol{\tilde{\psi}}| \boldsymbol{\tilde{\psi}}\rangle = \langle
\boldsymbol{\psi} | \boldsymbol{\psi} \rangle$. Hence,
using \eqref{eq:sun} leads to:
\begin{equation}
\langle \boldsymbol{\tilde{\psi}} | \boldsymbol{\tilde{\psi}} \rangle = \psi_{\alpha}
(U^{\dag})^{\alpha}_{\beta} U^{\beta}_{\gamma} \psi^{\gamma} = \psi_{\alpha}\psi^{\alpha} = \langle \psi| \psi \rangle
\end{equation}
which, for  arbitrary  $\boldsymbol{\psi}$,  is only possible provided
 $(U^{\dag})^{\alpha}_{\beta} U^{\beta}_{\gamma} = \delta^{\alpha}_{\gamma}$, that is,
in matrix notation:
\begin{equation}
U^{\dag} U = U U^{\dag} = \boldsymbol{1}, \label{eq:unity}
\end{equation}
where $\boldsymbol{1}$ is the unit matrix.  Hence,  $U^{-1} = U^{\dag}$, or,
in other words,  $U$ is a \emph{unitary} matrix. Furthermore, it also
 follows that  $\mathrm{det} \left[ U^{\dag} U \right]
 = \mathrm{det} \, \boldsymbol{1} = 1$, which
 implies that $\left| \mathrm{det} (U)\right|^2 = 1$. If
 $U$ belongs to SU$(N)$, then $ \mathrm{det} (U) = \epsilon_{\alpha_1 \cdots
 \alpha_N} U^{\alpha_1}_{1} \cdots U^{\alpha_N}_N = 1$, which generalizes
 to $\epsilon_{\alpha_1 \cdots
\alpha_N} U^{\alpha_1}_{\beta_1} \cdots U^{\alpha_N}_{\beta_N} = \epsilon_{\beta_1\cdots \beta_N}$.
The vector and its dual define the two fundamental irreducible representations
of SU$(N)$, which are denoted as $N$ and $\bar{N}$, respectively. We can consider tensors  with upper and lower indices which
transform as products of these two fundamental representations. For instance,
$\varphi^{\alpha\beta}$ belongs to the $N\times N$ tensor which transforms as $\psi^{\alpha}\chi^{\beta}$. The tensor $\varphi^{\alpha}_{\beta}$ belongs to
$N \otimes \bar{N}$  representation
transforming as $\psi^{\alpha}\chi_{\beta}$. It is worth noting that
the transformation properties of the tensors respect the permutation
symmetries of their indices. Thus, for $\varphi^{(\alpha\beta)} = \varphi^{\alpha\beta} = \varphi^{\beta\alpha}$ ($\varphi^{[\alpha\beta]} = \varphi^{\alpha\beta} = -\varphi^{\beta\alpha}$) a(n) (anti-)symmetric tensor, the transformed tensor
$\tilde{\varphi}^{\alpha\beta} = U^{\alpha}_{\gamma} U^{\beta}_{\delta} \varphi^{\gamma\delta}$ is also (anti-)symmetric. Hence, since the tensor $\varphi^{\alpha\beta}
= \varphi^{[\alpha\beta]} + \varphi^{(\alpha\beta)}$,
where $(\ldots)$  stands for symmetrization
of the indices and $[\ldots]$ for symmetrization, we have that
the representation $N\otimes N$ is reduced to $N(N-1)/2  \oplus N(N+1)/2$. Furthermore, an SU$(N)$ transformation respects the trace of a tensor  (the latter being understood as the result
of contracting an upper and a lower index). Hence, for instance,
$\varphi^{\alpha}_{\beta} = \frac{1}{N}\varphi^{\alpha}_{\alpha} \delta^{\alpha}_{\beta} +  \left(\varphi^{\alpha}_{\beta} - \frac{1}{N} \varphi^{\alpha}_{\alpha}\right)$, that is, $N\otimes\bar{N} = 1 \oplus N^2-1$.

Finally, let us consider the linear transformations  in the neighborhood
of the unit element of the group  (i.e. $\boldsymbol{1}$). For $N = 2$,
$\mathrm{SU}(2)\simeq O(3)$, the rotation group, and for this group
it is known that any finite  rotation  can be obtained as the product of an infinite set
of infinitesimal rotations. The latter differ from unity $\boldsymbol{1}$ by an infinitesimal amount, i.e. $U = \boldsymbol{1} + i \epsilon T$, where $\epsilon \ll 1$ is a real parameter and $T$ is a matrix whose properties we determine in what follows. From
\eqref{eq:unity} it follows that  $U^{\dag} U = (\boldsymbol{1} - i  \epsilon T^{\dag}) (\mathbf{1} + i \epsilon T) = \boldsymbol{1} - i\epsilon (T^{\dag} - T) + O(\epsilon^2) = \boldsymbol{1}$,  that is,
\begin{equation}
T^{\dag} = T.   \label{eq:hermit}
\end{equation}
Moreover, the unit determinant condition requires that
$1 = \mathrm{det} \: U = \mathrm{det} \: \left(\mathbf{1} + i \epsilon T \right)  = \mathrm{tr} \exp\left[ \ln(\mathbf{1} + i \epsilon T) \right] = 1+ i \epsilon\,  \mathrm{tr}\: T$,
where we have employed the identity $\mathrm{det} \: A = \mathrm{tr} \exp \left[ \ln  \: A \right]$. Therefore,
$\mathrm{tr} \: T = 0$, thus, the $N\times N$ matrices $T$ are hermitian (see ~\eqref{eq:hermit}) and
\emph{traceless}.  When expressed in terms of the matrix components, \eqref{eq:hermit} reads $(T^{\beta}_{\alpha})^* =
T^{\alpha}_{\beta}$. In other words, the diagonal elements of $T$ are real, and the $N(N-1)/2$
upper  and lower diagonal  are the complex conjugate to each other. Hence, it follows that $T$ depends only on $2 \times N (N-1)/2 + N = N^2$ real parameters. The traceless condition imposes a further constraint, which yields $N^2 - 1$ for the number of independent
$T$ matrices, which are denoted as $T^a$, with $a = 1, \ldots, N^2 - 1$.  Thus,  a   general infinitesimal $\mathrm{SU}(N)$
transformation can be written as $U = 1 + i  \sum_{a}\epsilon_a T^a$, where $\epsilon_a \ll 1$ are $N^2 - 1$ \emph{real}
numbers. For $N = 2$, there are $2^2 - 1 = 3$ matrices  proportional to the Pauli matrices $T^a = \frac{1}{2}\sigma^a$,
$a = x, y, z$. The latter satisfy the angular momentum algebra $\left[ T^a, T^b \right] = i \epsilon^{ab}_c T^c$, where
$\epsilon^{ab}_c$ is the fully anti-symmetric Levi-Civita symbol. This is an example of a \emph{Lie} algebra. For $\mathrm{SU}(N > 2)$, the Lie algebra is characterized by a set of structure
constants $f^{ab}_c$ different from $\epsilon^{ab}_c$:
\begin{equation}
\left[ T^a, T^b \right] = i f^{ab}_c T^c,
\end{equation}
The $N^2-1$, $N\times N$ traceless hermitian matrices, $T^a$, are the generators of the Lie algebra. Furthermore,  they also provide a basis for the linear space of $N\times N$ \emph{traceless} hermitian matrices.   Among them, we can distinguish
$N-1$ that are diagonal (like $T^3 = \sigma^z/2$ for $\mathrm{SU}(2)$), which form a set known as the Cartan basis. A representation for Cartan basis matrices is $T_3 = \frac{1}{2} \mathrm{diag}(1, -1, 0, \ldots, 0),\, T_8 = \frac{1}{\sqrt{12}}  \mathrm{diag}(1, 1, -2, 0,\ldots, 0),
\ldots, \, T^{r^2-1} = \frac{1}{\sqrt{2(2r-1)}}(1, 1, 1, \ldots,  -r, \ldots, 0)$, for $r = 2, \ldots, N$.
The other matrices are chosen hermitian and  non-diagonal and contain a single non-vanishing element equal to either $1/\sqrt{2}$ or $i/\sqrt{2}$.
This basis is conveniently normalized so that $\mathrm{Tr}\:  T^a T^b = \frac{1}{2} \delta^{ab}$. Another convenient basis for $\mathrm{U}(3) = U(1)\times SU(3)$ is provided by the projection
operators $X^{\alpha}_{\beta} = |\alpha\rangle \langle \beta|$, where $a,b = 1, \ldots, N$. In this basis, the Lie algebra takes a very simple form:
\begin{equation}
\left[ X^{\alpha}_{\beta},X^{\gamma}_{\delta} \right] = \delta^{\beta}_{\gamma} X^{\alpha}_{\delta} - \delta^{\alpha}_{\delta} X^{\gamma}_{\beta}.
\end{equation}
 Furthermore, $n =  X^{\alpha}_{\alpha}$ commutes with
all the generators $X^{\alpha}_{\beta}$, and corresponds to the generator
of the $\mathrm{U}(1)$ subgroup in $\mathrm{U}(3) = \mathrm{U}(1) \times \mathrm{SU}(3)$. Note that the non-diagonal  generators ($\alpha\neq \beta$) are not
hermitian, whereas the diagonal ones are not traceless.
However, this basis has the advantage that
it can be readily represented in second quantization: Let $c_{\alpha}$ transforms according to the $N$
irrep, and $c^{\dag}_{\alpha}$ transform according to $\bar{N}$,
then $X^{\alpha}_{\beta} = c^{\dag}_{\beta} c^{\alpha}$,
provided $n = c^{\dag}_{\alpha} c^{\alpha} = 1$.

\section{Fermi Liquid Parameters}\label{app:landau}

We can exploit the SU$(N)$ symmetry and write the Landau QP occupation and the Landau function as follows~\cite{Cazalilla2009}:
\begin{align}
\delta n^{\alpha}_{\beta}(\boldsymbol{p}) &= \frac{1}{N}\delta\rho_c(\boldsymbol{p}) \delta^{\alpha}_{\beta} +
\sum_{a=1}^{N^2-1} m_a(\boldsymbol{p}) \left(T^a\right)^{\alpha}_{\beta},\\
 f_{\alpha \gamma}^{\beta\delta}(\boldsymbol{p},\boldsymbol{p}^{\prime})
 &= f^{\rho}(\boldsymbol{p},\boldsymbol{p}^{\prime}) \delta^{\alpha}_{\beta} \delta^{\gamma}_{\beta} + 2 f^{m} (\boldsymbol{p},\boldsymbol{p}^{\prime})
 \sum_{a=1}^{N^2-1}  \left(T^a\right)^{\alpha}_{\beta}  \left(T^a \right)^{\gamma}_{\delta}, \label{eq:landaufun}
\end{align}
where  we have exploited the fact that $\delta   n^{\alpha}_{\beta}(\boldsymbol{p})$ is a $N\times N$ (hermitian) density matrix  which
can be split into a trace [$\delta\rho_c(\boldsymbol{p}) =  \delta   n^{\alpha}_{\alpha}(\boldsymbol{p})$], which describes fluctuations in the total
QP number,  and a traceless part. The latter can be conveniently expanded in terms of the generators of the SU$(N)$ algebra (see  Appendix~\ref{app:sun}) and describes the nuclear spin fluctuations. In group theoretic language, $\delta   n^{\alpha}_{\beta}(\boldsymbol{p})$ is a rank$-2$ tensor in the (reducible) $N\otimes \bar{N} = 1 \oplus N^2 -1$ representation (see Appendix~\ref{app:sun} for definitions).  Likewise, the fourth rank tensor of Landau functions belongs to the (reducible) representation
$N\otimes\bar{N}\otimes N \otimes \bar{N} = 1 \oplus 1$ + non-scalar representations, and therefore it can be parametrized in terms of two scalar functions as in Eq.~\eqref{eq:landaufun}. Because the QP are only well-defined excitations in the neighborhood of the FS (otherwise the strongly scatter each other),
 for $|\boldsymbol{p}| = |\boldsymbol{p}^{\prime}| \approx p_F$, rotational invariance requires that the Landau functions depend only on  $\cos \theta = \boldsymbol{p}\cdot \boldsymbol{p}^{\prime}/p^2_F$.  Thus, it is conventional~\cite{pethick_baym_book} to express the Landau functions $f^{\rho}(\cos\theta)$ and $f^m(\cos \theta)$ in terms the dimensionless Landau parameters $F^{\rho,m}_L$:
\begin{align}
f^{\rho}(\cos \theta) = \left[N\times N^0(\mu)\right]^{-1}
\sum_{L=0}^{+\infty}  F^{\rho,m}_L P_L(\cos \theta), \\
f^{m}(\cos \theta) = \left[N^0(\mu)\right]^{-1}
\sum_{L=0}^{+\infty}  F^{m}_L P_L(\cos \theta),
\end{align}
where $N^0(\mu) = p_F m^*/(２\pi^2 \hbar^2)$ is the QP density of states per spin
at the Fermi level and $P_L(\cos \theta)$ are the Legendre polynomials of order $L$.

\section{Nambu-Goldstone modes of SU$(N)$ superfluids}\label{app:ng}

To illustrate this point, let us consider the SU$(N=3)$ case~\cite{Honerkamp2004b,He2006,Yip2011}.
The order parameter is a rank-$2$ tensor that transforms
according to the
 $\bar{3}$ irreducible representation of SU$(3)$  (recall that $3\otimes 3 =  \bar{3} \oplus 6$, where $\bar{3}$ is the anti-symmetric representation, see Appendix~\ref{app:sun}). This is
made apparent by introducing the (complex) spinor $\boldsymbol{\Phi}$ whose components are $\Phi_{\alpha} = \epsilon_{\alpha\beta\gamma} \Delta^{\beta\gamma}$, where
$\epsilon_{\alpha\beta\gamma}$ is the fully anti-symmetric
Levi-Civita symbol. Applying Youla's decomposition,
we can use a Gauge for which $\Delta^{12} = \phi_0 \neq \ 0$ and the other components are zero, which means that we can always choose $\Phi = (0, 0, \phi_0)$. Consequently, the little group that is, the symmetry group that leaves the
order parameter invariant is SU$(2) \times$ U$(1)$, where
SU$(2)$ acts upon the first two components of $\boldsymbol{\Phi}$ whereas the U$(1)$ group acts on the phase of the third (non-zero) component.
Thus, the little group contains $3+ 1 = 4$ generators,  meaning in  U$(3) =$ U$(1)\times$SU$(3)$ ($9$ generators) there are $9 - 4 = 5$ broken-symmetry generators~\cite{Honerkamp2004b}.  However, in non-relativistic systems, the number of  NG modes is not equal to the number of broken symmetry generators (see e.g.
Ref.~\cite{Murayama2012} and references therein).  Qualitatively, this can be
understood by writing down an effective Lagrangian
for the order parameter spinor field $\bold{\Phi}(\mathbf{r})$.
Besides the U$(3)$ symmetry, the latter is constrained by
space rotation invariance, which leads to
\begin{equation}
\mathcal{L} = i  \boldsymbol{\Phi}^{\dag}(\boldsymbol{r},t) \partial_t \boldsymbol{\Phi}(\boldsymbol{r},t) - \frac{K_1}{2}
\nabla \boldsymbol{\Phi}^{\dag}(\boldsymbol{r},t) \cdot
\nabla \boldsymbol{\Phi}^{\dag}(\boldsymbol{r},t) - V(\boldsymbol{\Phi}^{\dag}\boldsymbol{\Phi}) + \cdots
\label{eq:efflag}
\end{equation}
where $K_1$  is a constant, the potential $V(\boldsymbol{\Phi}^{\dag}\boldsymbol{\Phi})$ has a minimum for $\boldsymbol{\Phi}^{\dag}\boldsymbol{\Phi} = \phi^2_0$, e.g. $V = \frac{\lambda}{2}
\left(\boldsymbol{\Phi}^{\dag} \Phi - \phi_0^2 \right)^2$ (we take $\phi_0$ real without loss of generality), and the dots stand for higher order gradient terms. Note that in a relativistic (i.e. Lorentz-invariant) or in a particle-hole symmetric  theory,
the first term in the right-hand side of Eq.~\eqref{eq:efflag} would be forbidden and should be replaced by
$\sim \partial_t \boldsymbol{\Phi}^{\dag} \partial_t \boldsymbol{\Phi}$. For such theories, the number of NG modes
equals the number of broken symmetry generators~\cite{Murayama2012}. However,
in the non-relativistic case, as  we shall see next, this term
is responsible for a dramatic change in the number and long-wave length dispersion of the NG modes.  If we parameterize the small fluctuations about the minimum as $\Phi(\boldsymbol{r},t) = \left(\phi_1(\boldsymbol{r},t), \phi_2(\boldsymbol{r},t),
\left[\phi_0 + \delta\rho_3(\boldsymbol{r},t)\right] e^{i\theta(\mathbf{r},t)}\right)$, it can be seen that the linearized equations of motion for $\theta(\boldsymbol{r}, t) $ and $\phi_{1,2}(\boldsymbol{r},t)$ imply that
the phase ($\theta$) NG mode has linear dispersion, i.e $\omega
\propto q$ for $q\to 0$. However, the $\phi_{1,2}$ NG modes disperse quadratically, i.e.
 $\omega\propto q^2$ as  $q \to 0$. Furthermore, the number of NG
modes is three, which is different from the number of Broken symmetry generators ($=5$) because, upon quantization, the fields $\phi^*_{1}, \phi^*_{2}$ and $\phi_{1},\phi_2$ cannot be treated as independent degrees of freedom as they correspond to the creation and annihilation of the same eigenmode.\cite{He2006,Yip2011}. Another lesson from this example is that quadratic modes correspond to
fluctuations in the paring function of the two paired components with unpaired one, i.e.  $\Phi_{1}= \Delta_{23}$ and $\Phi_2 = \Delta_{13}$. This is because, in the linearized
equations that follow from \eqref{eq:efflag}, $\phi_{1},\phi_2$ are
not coupled to each other and to $\phi_3$. However,
$\delta \phi_3 = \phi_3 -\phi_0 \sim \delta \rho_3 \: e^{i \theta}$, and $\delta \phi^*_3 \sim  \delta \rho_3 \: e^{-i \theta}$ are coupled, which leads to the linear dispersion for $\theta$.
This observation also generalizes to larger odd values of $N=5, 7, \ldots$, implying that for $N = 5$ there are four quadratic NG modes, etc. The quadratic coupling of the NG modes involving the unpaired component can also be explained using Gauge invariance arguments~\cite{Yip2011}. For even values of $N$,
the NG modes disperse linearly at small
$q$~\cite{Yip2011}.

\bibliography{./molref4}
\end{document}